\documentclass[10pt,journal,compsoc]{IEEEtran}

\ifCLASSOPTIONcompsoc
  \usepackage[nocompress]{cite}
\else
  \usepackage{cite}
\fi

\usepackage{cite}
\usepackage{times}
\usepackage{soul}
\usepackage{url}
\usepackage[utf8]{inputenc}
\usepackage[small]{caption}
\usepackage{graphicx}
\usepackage{booktabs}
\usepackage{amsmath}
\usepackage{amsthm}
\usepackage{mathrsfs}
\usepackage{verbatim}
\usepackage{multirow}
\usepackage{colortbl}
\usepackage{epstopdf}
\usepackage{diagbox}
\usepackage{multirow}
\usepackage{booktabs}
\usepackage[ruled]{algorithm2e}
\usepackage{amsfonts}
\usepackage{float}
\usepackage{stackengine}
\usepackage[table,xcdraw]{xcolor}
\usepackage{soul} 
\usepackage{color, xcolor}
\usepackage{tcolorbox}
\usepackage{pifont}
\usepackage{tabularx}
\usepackage[breaklinks]{hyperref}

\newcommand{\toolname}{\texttt{ANGEL}\xspace}

\newcommand{\eg}{\textit{e}.\textit{g}.,\xspace}
\newcommand{\ie}{i.e.,\xspace}

\newcommand*{\mycode}{\fontfamily{lmtt}\selectfont}

\begin{document}
\title{Keep It Simple: Towards Accurate Vulnerability Detection for Large Code Graphs}

\author{
        Xin~Peng,
        Shangwen~Wang,
        Yihao~Qin,
        Bo~Lin,
        Liqian~Chen,
        Jieren~Cheng,
        and~Xiaoguang~Mao
\IEEEcompsocitemizethanks{\IEEEcompsocthanksitem Xin Peng, Shangwen Wang, Yihao Qin, Bo Lin, Liqian Chen, and Xiaoguang Mao are with the Key Laboratory of Software Engineering for Complex Systems, College of Computer Science and Technology, National University of Defense Technology, Changsha, China. \\ E-mails: xinpeng@nudt.edu.cn, wangshangwen13@nudt.edu.cn, yihaoqin@nudt.edu.cn, linbo19@nudt.edu.cn, lqchen@nudt.edu.cn, and xgmao@nudt.edu.cn \protect

 \IEEEcompsocthanksitem Jieren Cheng is with the Hainan University, Haikou, China. \\ E-mails: cjr22@163.com \protect
 \IEEEcompsocthanksitem Shangwen Wang is the corresponding author.
 }
}

\markboth{ }%
{Xin Peng \MakeLowercase{\textit{et al.}}: }

\IEEEtitleabstractindextext{%
\begin{abstract} 
Software vulnerability detection is crucial for high-quality software development. Recently, some studies utilizing Graph Neural Networks (GNNs) to learn the graph representation of code in vulnerability detection tasks have achieved remarkable success. However, existing graph-based approaches mainly face two limitations that prevent them from generalizing well to large code graphs: (1) the interference of noise information in the code graph; (2) the difficulty in capturing long-distance dependencies within the graph. 

To mitigate these problems, we propose a novel vulnerability detection method, \toolname, whose novelty mainly embodies the hierarchical graph refinement and context-aware graph representation learning. 
The former hierarchically filters redundant information in the code graph, thereby reducing the size of the graph, while the latter collaboratively employs the Graph Transformer and GNN to learn code graph representations from both the global and local perspectives, thus capturing long-distance dependencies. 
Extensive experiments demonstrate promising results on three widely used benchmark datasets: our method significantly outperforms several other baselines in terms of the accuracy and F1 score. Particularly, in large code graphs, \toolname achieves an improvement in accuracy of 34.27\%-161.93\% compared to the state-of-the-art method, AMPLE. Such results demonstrate the effectiveness of \toolname in vulnerability detection tasks.
\end{abstract}

}

\maketitle

\IEEEdisplaynontitleabstractindextext
\IEEEpeerreviewmaketitle

\section{Introduction}
\label{sec:Introduction}
\IEEEPARstart{S}{oftware} vulnerabilities are specific defects and weaknesses within software components. Attackers can exploit these vulnerabilities to carry out malicious actions, endangering the lives of human societies \cite{CSUR2017survey}. For example, the notorious ``WannaCry'' ransomware exploited operating system vulnerabilities to affect hundreds of thousands of computers worldwide, causing serious economic losses and social impacts \cite{ICMLA2017wannacry}. Therefore, the research and development of software vulnerability detection techniques have attracted widespread attentions from researchers~\cite{IEEE2020survey, wang2020contractward, TDSC2021vuldeelocator, ICSE2022vulcnn, lin2023cct5,zhang2022reentrancy,wu2021peculiar}.

Traditional vulnerability detection techniques primarily relied on manual auditing and expert experience, which were labor-intensive, inefficient, and required comprehensive domain knowledge \cite{shahriar2012mitigating}. In addition, there were some other studies exploring to utilize static code analysis and dynamic behavior analysis, but these were still cumbersome and prone to miss concealed vulnerabilities \cite{ACM2007practical, ICDS2009fitness, USENIX2015under}. With the rapid development of deep learning technique, the automated detection of complex vulnerabilities has become a research hotpot \cite{TDSC2019mVulDeePecker, IJCAI2021smart, wang2023pre}. These methods generally rely on large amounts of annotated data to train deep learning models, thereby improving the accuracy and efficiency of vulnerability detection. Existing deep learning (DL)-based software vulnerability detection methods can generally be classified into two categories: token-representation-based and graph-representation-based. The token-based methods consider code as a natural language sequence, parsing the source code into a series of tokens, and then converting them into vectors for predictions \cite{Arxiv2018Vuldeepecker,TDSC2021Sysevr, ACSAC2022transformer}. However, unlike natural languages, source code contains rich structure and semantic information reflected by the abstract syntax tree (AST) and data/control flow relations.
Therefore, in recent years, methods based on graph representation achieve more advanced detection performances \cite{IST2021bgnn4vd, PACT2020Graph, CN2024DA-GNN, ICSE2024coca}. Typically, these methods use program analysis tools to construct various code structure graphs, such as control flow graph (CFG) and data flow graph (DFG), to represent complex relationships in the program \cite{TSE2023cpvd}. Subsequently, to effectively learn the information in the graphs, Graph Neural Networks (GNNs) are widely adopted by most advanced vulnerability detectors \cite{ICSE2022regvd} . For example, Devign \cite{Nips2019Devign} proposed a joint graph structure that includes AST, CFG, DFG, and natural code sequences, and used a gated graph neural network (GGNN) \cite{Arxiv2015GGNN} to build a vulnerability detection model. Additionally, there are methods that use manually predefined rules to construct novel code structure graphs, thereby enriching the semantic information of the constructed graph structure \cite{ICSE2024SCVHunter, TSE2023MAGNET}.

\begin{figure*}[!t]
\centering
\includegraphics[width=1.0\textwidth]{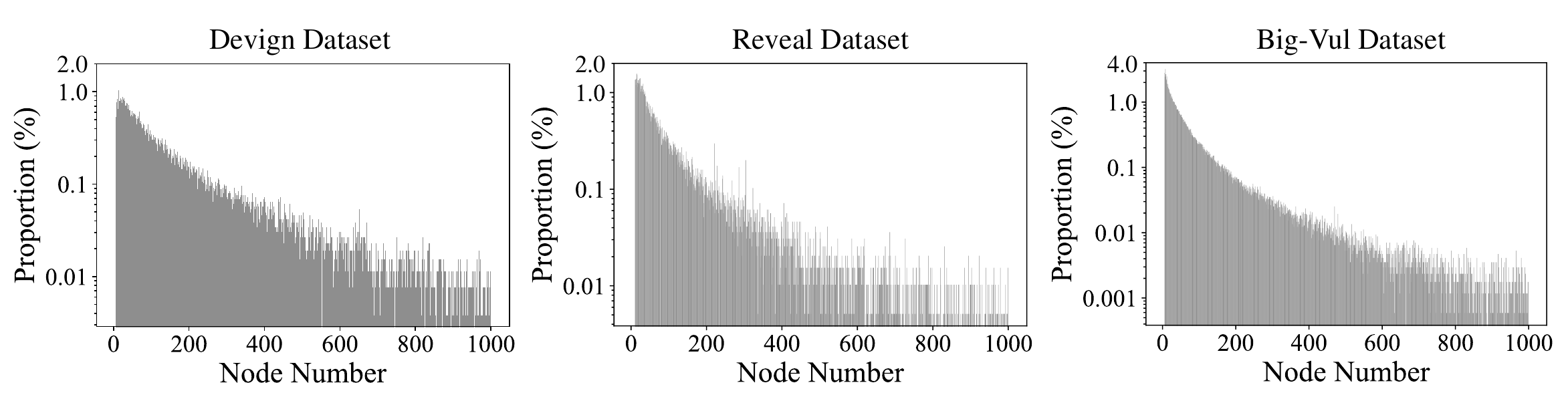}
\caption{Statistics on the distribution of graph sizes in the Devign \cite{Nips2019Devign}, Reveal \cite{TSE2021Reveal} and Big-Vul \cite{MSR2020BigVul} datasets.}
\label{NodeNum}
\end{figure*}

\begin{figure*}[!t]
\centering
\includegraphics[width=1.0\textwidth]{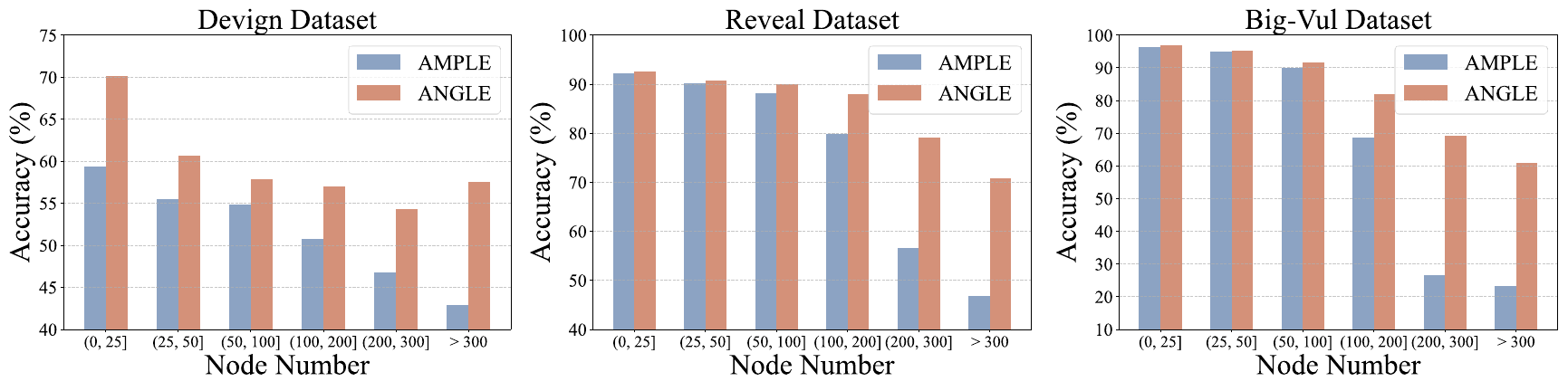}
\caption{Accuracy with different number of nodes in the Devign \cite{Nips2019Devign}, Reveal \cite{TSE2021Reveal} and Big-Vul \cite{MSR2020BigVul} datasets.}
\label{NodeAcc}
\end{figure*}

Although current graph-based methods exhibit promising results, one bottleneck that arises is their relatively poor performance when dealing with large code graphs. Specifically, Fig. \ref{NodeNum} shows the distribution of node numbers in the code property graphs \cite{ISP2014cpg} of different vulnerable programs from the widely-used Devign \cite{Nips2019Devign}, Reveal \cite{TSE2021Reveal} and Big-Vul \cite{MSR2020BigVul} datasets.
We note that the number of nodes generally exhibits a long-tail distribution, and such a phenomenon can be observed from all the three datasets, which indicates that this is a universal characteristic for real-world vulnerabilities.
Although the majorities are in the (0, 100] interval, there are still considerable samples with more than 100 nodes, and they are unevenly distributed across various intervals with the largest code graph containing more than 1K nodes (due to the space constraint, the maximum value is set to 1K in the figure). This disparity in distribution forces vulnerability detectors to bias towards fitting graphs with fewer nodes, leading to a performance bottleneck where large graphs are difficult to be precisely predicted \cite{ICML2021LocalSG, WWW2022SOLT-GNN, NIPS2022GAMPNN, ICML2023WBM}. 
As shown in Fig. \ref{NodeAcc}, we demonstrate the accuracy of AMPLE \cite{ICSE2023AMPLE}, the state-of-the-art graph-based vulnerability detectors, under different node count partitions on the three datasets. Taking the Reveal dataset as an example, AMPLE achieves an accuracy of 92.25\% in the (0, 25] interval, while its effectiveness significantly declines when the number of node exceeds 100 (\eg its accuracy is only 46.85\% when the number of nodes is larger than 300). Such results demonstrate that the performances of state-of-the-art vulnerability detection techniques severely decreases with the scales of code graphs increase.

\begin{figure*}[!t]
\centering
\includegraphics[width=1.0\textwidth]{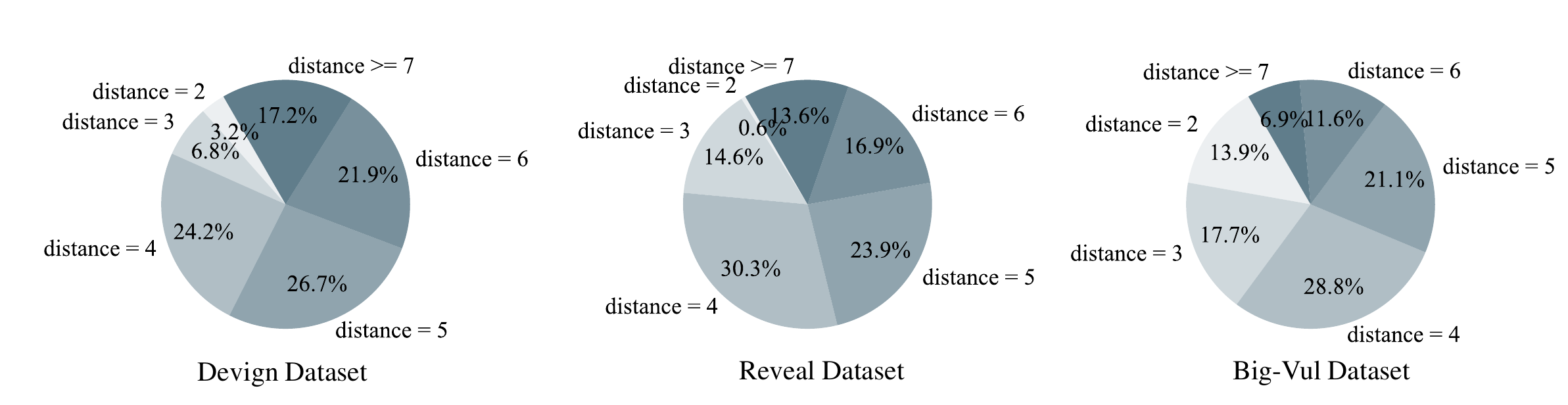}
\caption{Statistics of longest distance in the Devign \cite{Nips2019Devign}, Reveal \cite{TSE2021Reveal} and Big-Vul \cite{MSR2020BigVul} datasets.}
\label{NodeDia}
\end{figure*}

Upon reviewing the working mechanisms of existing approaches, their ineffectiveness on large code graphs could mainly be attributed to the following two reasons. \ding{192} {\bf The large amount of noisy information in the graph could affect the prediction.} 
Existing methods usually use global pooling after the GNN layers to convert the graph into a vector representation \cite{Nips2019Devign, FSE2021IVDetect, ICSE2023AMPLE}. In such a process, information from all the nodes in the graph is utilized to make the final decision. However, there might be only a small number of key nodes that incur the vulnerability while the other nodes are not directly related to the vulnerability \cite{IJCAI2019VulSniper, MSR2022linevd}. In such cases, the prediction of existing approaches could be influenced by the large amount of vulnerability-irrelevant noisy information, which correspondingly leads to inaccurate results.
Although AMPLE incorporates a graph simplification technique \cite{ICSE2023AMPLE}, this process relies on pre-defined heuristics and typically results in the reduction of only a limited number of nodes in the graph. This limitation could potentially explain the inefficacy of AMPLE on large code graphs, as will be demonstrated in Section~\ref{sec:motivation}.
\ding{193} {\bf The long-distance dependencies in the graph could be neglected.} 
By utilizing standard GNN layers, existing methods only gather information from neighboring nodes for generating the representation of a node \cite{TSE2021Reveal, ICSE2023AMPLE, TOSEM2021Deepwukong}. However, it is widely known that software code contains long-distance dependency relations. For instance, a variable defined at the beginning of a function might be used in the final return statement of that function.
Fig. \ref{NodeDia} displays the distribution of the longest distance between two nodes for the programs in the Devign \cite{Nips2019Devign}, Reveal \cite{TSE2021Reveal} and Big-Vul \cite{MSR2020BigVul} datasets. It can be observed that more than half of the programs contain a pair of nodes in their graphs whose distance exceeds or equal to four. 
As shown by existing studies, there could be challenges such as the over-smoothing and over-compression for GNNs to fit graphs whose longest path is greater than three \cite{ICLR2021over}, which may lead to the failure of GNNs of comprehensively understanding the graphs.
Consequently, critical dependency relations could be missed by existing approaches when detecting vulnerabilities in such programs.

In this paper, we propose a hierarchic\underline{A}l and co\underline{N}text-aware \underline{G}raph r\underline{E}presentation \underline{L}earning approach that aims at mitigating the ineffectiveness of existing vulnerability detection techniques on large code graphs. To overcome the aforementioned two limitations, \toolname contains two customized modules, \ie an importance-based graph simplification module and a global \& local feature aggregation module. 
After parsing the programs into graph structures, the importance-based graph simplification module hierarchically refines the graph, aiming at reducing the size of the graph so that vulnerability-irrelevant noisy information could be filtered. 
In contrast to AMPLE, our objective is to autonomously learn the characteristics of the pivotal nodes within the graph rather than depending on pre-established heuristics, which are recognized for their lack of flexibility.
To that end, Top-k pooling \cite{ICML2019TopK} is adopted to select critical nodes based on their importance scores, which are calculated by a learnable matrix. 
Note that the simplification process is self-adaptive: for each code graph, this phase terminates when the number of left nodes becomes lower than a pre-defined threshold. This means that there might be unnecessary to perform simplifications for small-scale code graphs, while large-scale graphs may iteratively undergo several numbers of pooling operations. 
This strategy enables us to retain the size of the simplified graph within a reasonable range. In contrast, the existing AMPLE typically only reduces a minimal portion of nodes in the original graph (less than 50\%) and thus the simplified graph can still be in large scale.
The number of filtered nodes is dynamically adjusted during different pooling layers. Specifically, in the first pooling step, \toolname only preserves 10\% of the total nodes for quickly removing noisy information; while in the last several pooling layers, 50\% of the input nodes are preserved for precisely identifying the critical information.
Taking the simplified graph as input, we utilize GNNs for feature extraction to generate a one-dimensional vector representation. Different from existing methods, \toolname borrows the weapon of Graph Transformer (GT) \cite{ICML2022GT1, NPS2022GT2}, a network that can model the semantic information of a graph from the holistic perspective, \ie the long-distance dependency relations could be captured by such a model. Notably, we are the first to integrate GNN and GT for the vulnerability detection task, leveraging their respective strengths to design a global \& local feature aggregation module. Due to the varying scale and complexity of each code graph, the generality of the state-of-the-art method AMPLE, which rely on GNN and Convolutional Neural Networks (CNN) with fixed kernel sizes, is inherently limited. In contrast, our feature aggregation module alternates between traditional GNN and GT layers, with residual connections enhancing its learning capacity. Specifically, GNN focuses on modeling the semantics of nodes from their adjacent neighbors, while GT leverages a multi-head self-attention mechanism to extract global features from the entire graph for representation generation. In this manner, both global and local context relationships in the graph are captured, which alleviates the issue of long-range dependencies and achieves a more accurate extraction of semantic information from the graphs. 

To evaluate the effectiveness of \toolname, we chose three commonly-used public datasets for experiments: Devign \cite{Nips2019Devign}, Reveal \cite{TSE2021Reveal}, and Big-Vul \cite{MSR2020BigVul}. We selected eight state-of-the-art learning-based vulnerability detection techniques as baselines for comparison, where two are based on tokens, and the remaining six are based on graphs. The experimental results demonstrate that \toolname achieves significant improvements compared to all the baselines. Specifically, \toolname achieves a 37.70\%-202.20\% improvements in terms of the F1 scores across the three datasets compared to the token-based methods, while such improvements become 11.55\%-164.01\% when it comes to the comparison with graph-based methods. 

This paper makes the following contributions: 
\begin{itemize}
    \item \textbf{Problem.} We reveal that due to the complexity of real-world projects, current graph-based vulnerability detection models fall short on detecting vulnerabilities in large code graphs, significantly limiting their generalization ability and performances.

    \item \textbf{Approach.} We propose a novel vulnerability detection approach called \toolname. This method reduces the sizes of the code graphs through a carefully designed graph simplification module and combines GNN with GT to learn diverse context features in the graph.

    \item \textbf{Evaluations.} Experiments on three public benchmark datasets indicate that our proposed \toolname outperforms the state-of-the-art baseline methods.
\end{itemize}

\section{Background \& Related Works}
\label{sec:background}
\subsection{Graph Neural Networks}
In the software engineering domain, graphs have been used to represent various structures of code, such as CFGs, PDGs, ASTs, and others \cite{wang2024fusing, wang2023two}.
These graph structures can reflect the structures of the program code, the dependencies between variables, as well as the processes of program execution \cite{ASE2020graph4se}. 
Due to their superior performance, GNNs have been adopted for learning graph data and have been widely applied in tasks such as code clone detection \cite{ICSE2022treecen} and vulnerability detection \cite{STVR2024tensor}. 
GNNs learn node representations or graph-level representations in an end-to-end manner. Specifically, each node communicates with its neighbors through a propagation operator, after which various aggregation functions (\eg sum, average, or max) are used to integrate information from the neighboring nodes \cite{TNNLS2020GNNSurvey}. Graph Convolutional Networks (GCN) \cite{peng2023dual} and Graph Attention Networks (GAT) \cite{peng2024multi} are classical examples of GNNs. GCN introduces the concept of convolution into graph neural networks. Based on GCN, GAT introduces an attention mechanism to provide different aggregation weights for the information from neighbor nodes. 
Previous graph-based vulnerability detection studies have employed several layers of traditional GNNs, such as GCN and GGNN, to capture local information in the code structure graphs.

Recently, researchers have observed the long-range dependency problem in GNNs, leading to the design of a specialized Transformer architecture for graph-structured data, known as the GT network \cite{Nips2024GT3}. 
GT dynamically computes interaction weights with all nodes through the attention mechanism, effectively capturing global information within the graph and adapting to complex graph structures. Additionally, with multi-head attention, GT can learn representations of nodes in different subspaces, enriching the feature representation of nodes. To fully capture different patterns of vulnerabilities, in this work, we combine the characteristics of GNN and GT, utilizing their complementarity to learn the code structure graphs from both global and local perspectives.

\subsection{Hierarchical Graph Representation Learning}
Hierarchical graph representation learning aims to mine finer-grained and important node information in graph structures through an iterative manner \cite{AAAI2020Asap}, where hierarchical pooling is usually utilized to generate representations for smaller subgraphs.

Typically, hierarchical pooling can be divided into two types: clustering-based pooling and node-dropping-based pooling \cite{IJCAI2023graphpool}.
The former transforms hierarchical learning into a clustering problem, in which nodes in the graph are partitioned into different clusters and new nodes will be generated based on the cluster centroids.
For example, DiffPool \cite{NIPS2018Diffpool} learns differentiable soft cluster assignments for nodes at every layer of GNNs, mapping related nodes to a set of clusters to form the input for the next layer of the network. 
One notable drawback of this type of method is that it usually has high time and space complexity.
The latter, \ie the node-dropping-based pooling, designs a learning scoring function to remove nodes with relatively low scores, thereby obtaining subgraphs composed by important nodes. 
For instance, Graph U-Nets \cite{ICML2019TopK} designs a similarity-based pooling operator to progressively learn fine-grained subgraphs, while GXN \cite{NIPS2020GXN} uses a local pooling operator designed based on maximizing mutual information.
Typically, in the code graphs, not all the nodes are directly related to the vulnerability. That is to say, a number of nodes could be unrelated to the vulnerability and may bring noisy for the detection results \cite{MSR2022linevd}. 
Therefore, in this work, we exploit hierarchical graph learning methods to retain important nodes in the code graphs based on node dropping, thereby filtering irrelevant information.

\subsection{DL-based Vulnerability Detection}
Learning-based approaches have been extensively applied to the task of vulnerability detection \cite{ASE2018contractfuzzer,IST2023HGIVul, CS2024VDTriplet, ISCE2024MCU}. Previous works have focused on representing source code as natural language sequences and applying sequence models to learn vector representations of vulnerability information \cite{TDSC2019multi-domain, IJCNN2019deep, TDCS2020cd}. Generally speaking, source code contains complex syntactic and semantic structures, which is why graph-based methods have achieved state-of-the-art performance \cite{ICSE2023AMPLE}. These methods typically use static analysis to transform the source code into a graph, and employ some form of word embedding to generate the initial attributes for nodes \cite{IJCAI2021smart}. Reveal \cite{TSE2021Reveal} utilizes a CPG to represent source code information, then employs a GGNN to capture local structural information in the code. DeepDFA \cite{ICSE2024deepDFA} designs a CFG with abstract dataflow embeddings to capture information such as API calls and operations, and then uses a GGNN to learn complex structural relations. Furthermore, some approaches design different network architectures based on various code representations to learn code representations. For example, Vul-SPG \cite{ISSRE2021Vu1SPG} proposes a sliced property graph and designs a triplet attention mechanism to enhance the capability of aggregating node information, thereby achieving better detection performance. Similarly, MAGNET \cite{TSE2023MAGNET} views the code structure as a heterogeneous graph and uses a graph neural network with an attention mechanism to learn vulnerability patterns. 

Nevertheless, these techniques can frequently fail when confronted with large code graphs and long-range dependencies \cite{ICSE2023AMPLE}. This is primarily because they indiscriminately assimilate knowledge from diverse source codes using conventional GNNs, which are ill-equipped to effectively address these specific challenges.
Our approach, while still grounded in graph structures, incorporates tailored solutions to tackle the two challenges, as will be detailed in Section~\ref{sec:architecture}.

\section{Motivating Example}
\label{sec:motivation}

\begin{figure}[!t]
\centering
\includegraphics[width=0.48\textwidth]{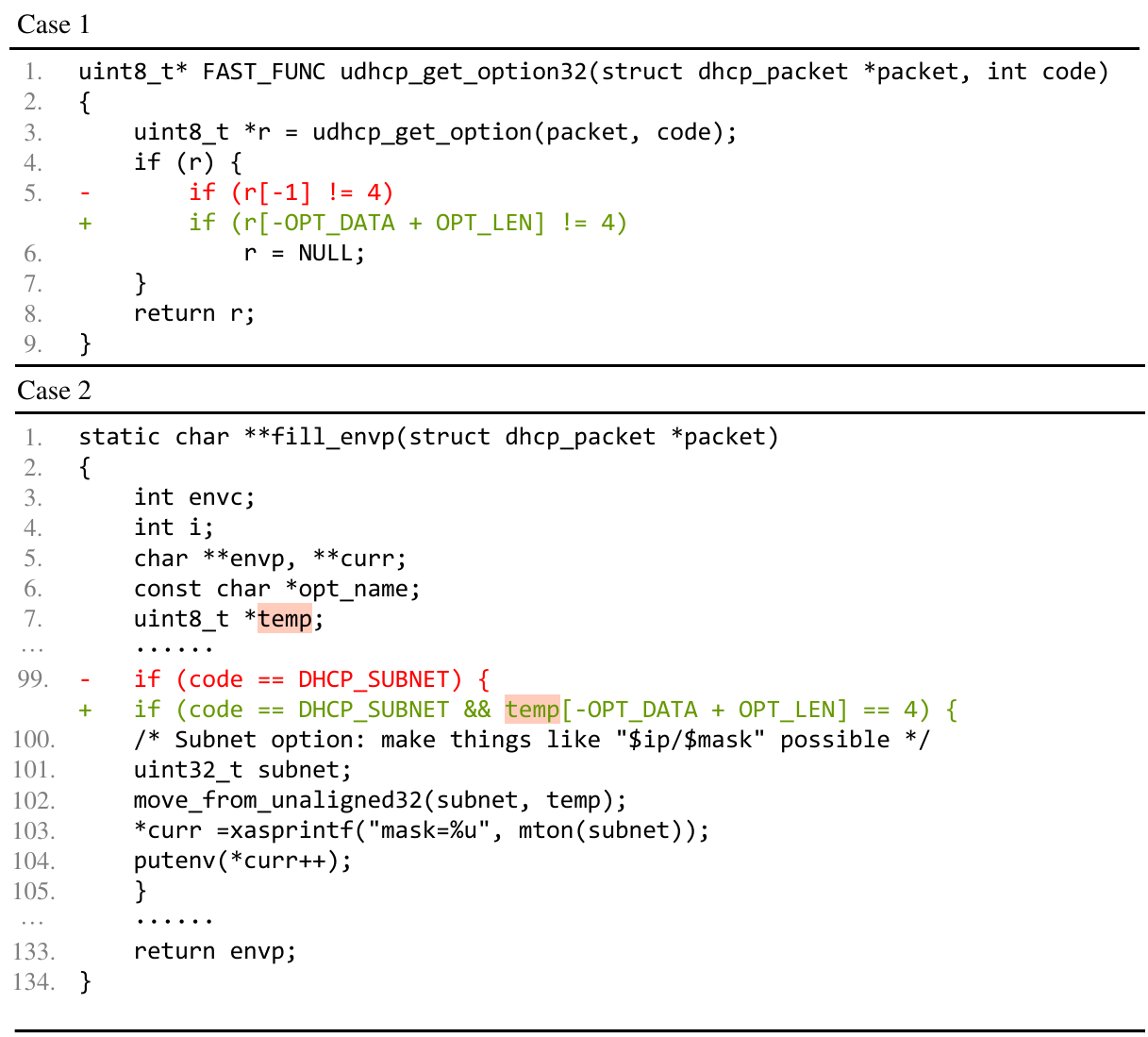}
\caption{Two different functions from the {\tt busybox} project have the same type of vulnerability (CWE-125). The red-colored code is the vulnerable code, and the green-colored code is the repaired code.}
\label{Moti}
\end{figure}

To demonstrate that the state-of-the-art graph-based vulnerability detection methods perform comparatively worse on large graphs, we select two detailed cases from the Big-Vul dataset as the example, both of which are with the out-of-bounds read vulnerability type (CWE-125).
These two examples are shown in Fig. \ref{Moti}, where brief excerpts of two different functions from the {\tt busybox} project are presented.\footnote{\url{https://git.busybox.net/busybox/}} These two examples have the same vulnerability type and developers even leave the same comment during the vulnerability fixing activities: ``\textit{when decoding DHCP\_SUBNET, ensure it is 4 bytes long}".
Nonetheless, one significant difference between these two functions is their code length. Specifically, Case 1 contains only 9 lines of code, while Case 2 possesses up to 134 lines of code.
As a result, when Case 2 is transformed into the code graph, it would contain a large number of nodes irrelevant to the vulnerability, which could influence the prediction of the state-of-the-art detection approaches and thus cast side effects on the detection result.\footnote{In this paper, vulnerability-relevant information denotes those that have data/control dependency with the variables from the vulnerable statements, while other entities in the code are denoted as vulnerability-irrelevant information.}

In fact, the state-of-the-art vulnerability detection approaches can indeed be misled by the redundancy information in the code graph.
Taking the state-of-the-art approach AMPLE \cite{ICSE2023AMPLE} as an example, it correctly detects the first case as a vulnerable function, however, it incorrectly predicts the second case as non-vulnerable.

We also recall that in the AMPLE study, the authors showcased AMPLE as the most adept existing method for managing large code graphs. Intuitively, if AMPLE encounters challenges in making accurate predictions, it is plausible that other state-of-the-art approaches may also encounter similar difficulties. Indeed, we have confirmed that other state-of-the-art approaches like Devign and Reveal also fail to predict the vulnerability in Case 2.

Besides the noisy information in the code graph, another challenge brought by the large-scale code snippet is the long-range dependencies in the graph which may not be easily captured by the existing approaches. 
Taking Case 2 as an example, the array related to the vulnerability, \ie {\mycode temp}, is first initialized in line 7, however, the buggy line that needs to check the content of {\mycode temp} occurs in line 99. Between these two statements, {\mycode temp} undergoes two assignments, two conditional expressions where it is used to judge if the condition is satisfied, and it is also used as an argument for one time.
Therefore, it is quite hard for existing approaches to comprehensively comprehend the data and control dependency relations of the array {\mycode temp}, and thus fail to recognize that the buggy line also needs to check the content of {\mycode temp}.

From the above examples, we find that state-of-the-art vulnerability detection approaches perform relatively poor on large code graphs, probably due to the following two reasons: (1) {\bf they could be misled by the vulnerability-irrelevant information in the large graphs}, and (2) {\bf they fall short on understanding the long-range dependencies in the large graphs}. 
Therefore, we are motivated to design a vulnerability detection method by designing strategies to address these two limitations respectively, thus making it able to generalize well to large code graphs.

\section{Approach}
\label{sec:architecture}

\begin{figure}[!t]
\centering
\includegraphics[width=0.48\textwidth]{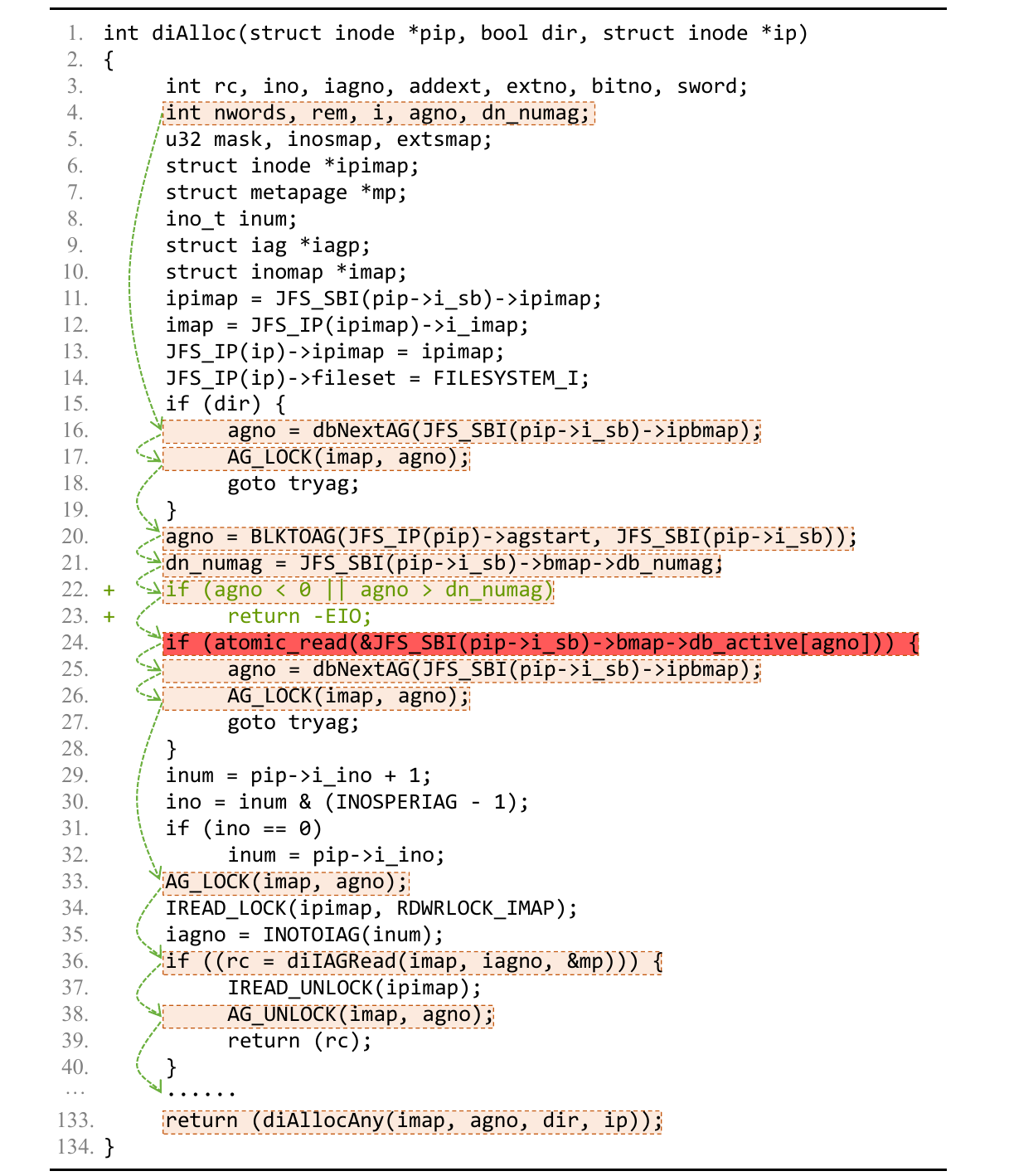}
\caption{An example of a vulnerability in the Linux kernel.}
\label{pivotalnode}
\end{figure}

\begin{figure*}[!t]
\centering
\includegraphics[width=\textwidth]{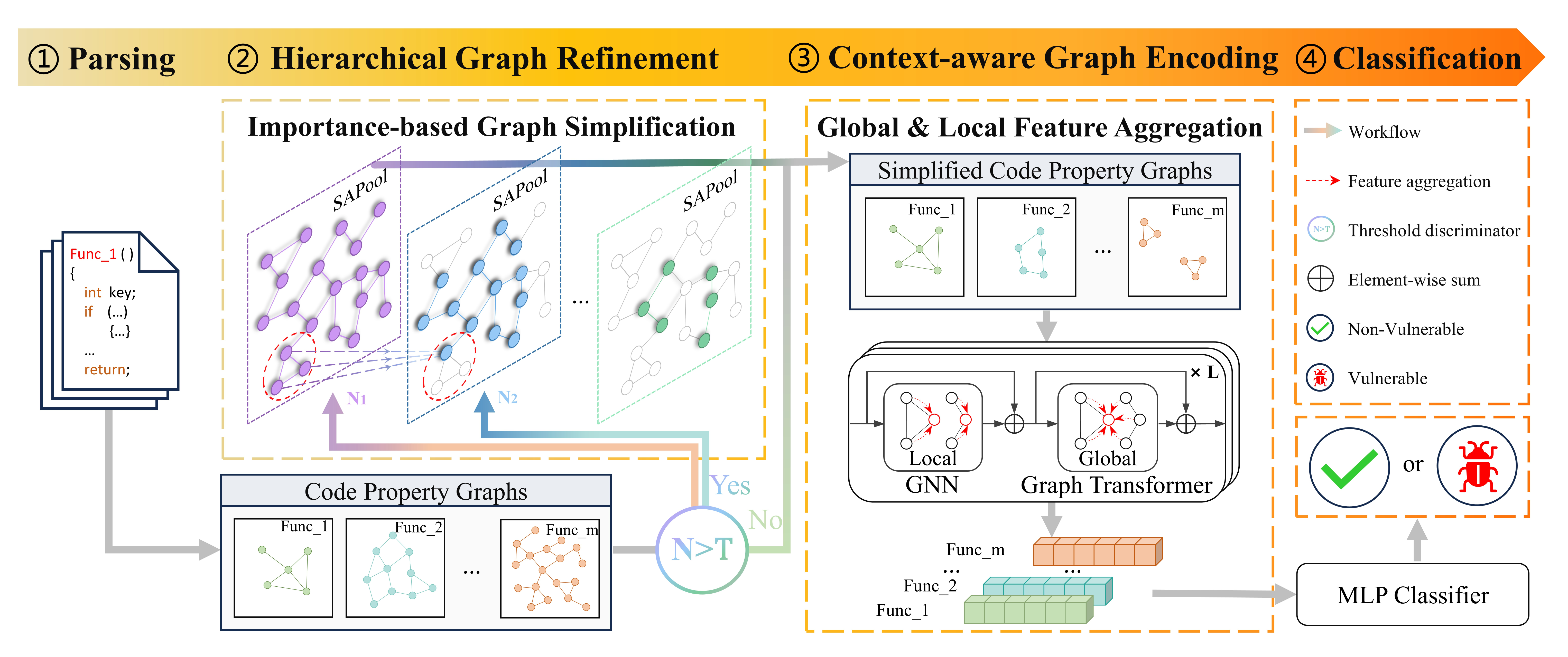}
\caption{The architecture of \toolname, which mainly contains two core module: (1) importance-based graph simplification module; (2) global \& local feature aggregation module. The four-stage paradigm (\ie parsing, hierarchical graph refinement, context-aware graph encoding, and classification) endows \toolname with enhanced effectiveness for vulnerability detection.}
\label{model}
\end{figure*}

\subsection{Overview}

In this work, we propose a vulnerability detection approach \toolname, aiming at addressing the challenges brought by large code graphs. 
The key novelty of \toolname lies in its (1) self-adaptive learning of important nodes in the graph to hierarchically reduce the graph size, and (2) collaborative utilization of GNN and GT to aggregate the features from both local and global contexts.
Generally, \toolname works in a four-stage paradigm that consists of parsing, hierarchical graph refinement, context-aware graph encoding, and classification. 
In the following, we first provide some basic definitions and then introduce each stage briefly.

\subsubsection{Symbol and Task Definition}
Given a set of graphs ${\mathcal{G} = \{G_1, G_2,...,G_M\}}$ associated with the corresponding labels ${\mathcal{Y} = \{y_1, y_2,...,y_M\}}$ and node numbers ${\{N_1, N_2,...,N_M\}}$. For each graph ${{G}_i = \{\mathcal{V}^{G_i}, \mathcal{E}^{G_i}, \mathbf{X}^{G_i}\}}$, the corresponding adjacency matrix is ${\mathbf{A} \in \mathbb{R}^{N \times N}}$, and the degree matrix is ${\mathbf{D} \in \mathbb{R}^{N \times N}}$. Here, ${\mathcal{V}}$ represents the set of nodes, ${\mathcal{E}}$ represents the set of edges, and ${\mathbf{X} \in \mathbb{R}^{N \times d}}$ denotes the raw feature matrix. The symmetric normalized adjacency matrix ${\hat{{\mathbf{A}}} \in \mathbb{R}^{ N \times N}}$ is calculated through ${{\mathbf{D}}^{-\frac{1}{2}}({\mathbf{A}}+{\mathbf{I}}) {\mathbf{D}}^{-\frac{1}{2}}}$, where ${\mathbf{I} \in \mathbb{R}^{N \times N}}$ is an identity matrix.

Given a vulnerable sample $F_v$, it typically consists of the vulnerability lines $f_l$, which explain how the vulnerability arises; the contextual statements $f_c$, which are associated with the vulnerability; and a significant number of statements $f_n$ that are not directly related to the vulnerability. The sample $f_v$ is formally defined as:
\begin{equation}
F_v = f_l \cup f_c \cup f_n
\end{equation}
Here, $f_l$ represents the core characteristics that expose the vulnerability, $f_c$ aids in understanding the conditions under which the vulnerability emerges, and $f_n$ may act as distracting information in the vulnerability detection task. We employ the analysis tool Joern \footnote{\url{https://joern.io/}} to transform $ F_v $ into a code property graph $ G_v = \{\mathcal{V}, \mathcal{E}, \mathbf{X}\}$. Accordingly, the sets of nodes corresponding to the code lines $ f_l $, $ f_c $, and $ f_n $ are defined as $ \mathcal{V}_l $, $ \mathcal{V}_c $, and $ \mathcal{V}_n $, respectively. Next, the edges present in the AST, CFG, and PDG are defined as $ \mathcal{E} $, which connect all nodes from the node sets $ \mathcal{V}_l $, $ \mathcal{V}_c $, and $ \mathcal{V}_n $. Each node in the set $\mathcal{V}$ possesses foundational information (such as node type and source code snippet) to represent the semantics of the node. We encode this information as vectors, which serve as the feature matrix $\mathbf{X}$ in the code property graph $G_v$. The attribute characteristics of each node can be formally expressed as $x$. Since focusing on the vulnerable lines $f_l$ and the associated contextual statements $f_c$ is essential for uncovering the root causes of vulnerabilities, we regard the corresponding nodes $ \mathcal{V}_l $ and $ \mathcal{V}_c $ as \textbf{key nodes}. The structural relationships between and within these statements are represented by the edges $\mathcal{E}_p$ in the code graph. Accordingly, the attribute information $ x $ contained within each node of $\mathcal{V}_l$ and $\mathcal{V}_c$, along with the edges $ \mathcal{E}_p $ connecting these nodes, together form the \textbf{characteristics of key nodes} in the code graph $G_v$. We use the real-world vulnerability CVE-2023-52805 \footnote{\url{https://nvd.nist.gov/vuln/detail/cve-2023-52805}} as an example. The green code snippet in Fig. \ref{pivotalnode} demonstrates the patch for this vulnerability, with red backgrounds indicating the vulnerable lines of code, and orange backgrounds representing the associated statements related to the vulnerability. In the unpatched code, the absence of boundary checks for the value of the variable {\tt agno} leads to the possibility of out-of-range values. To detect this vulnerability, it is crucial to focus on the statements involving the variables {\tt agno} and {\tt dn\_numag} (such as those related to computation logic and boundary checks). In this case, the nodes corresponding to the lines of code with red backgrounds are defined as $\mathcal{V}_l$, and those with orange backgrounds as $ \mathcal{V}_c$. The $ \mathcal{V}_l $ and $ \mathcal{V}_c$ nodes are considered pivotal nodes, as they carry the contextual information related to the vulnerabilities. The dependency relationships between these statements are implemented through the PDG, defined as $ \mathcal{E}_p $ (\ie the green arrows in Fig. \ref{pivotalnode}). By concentrating on these nodes, the learning-based model effectively captures essential patterns associated with potential vulnerabilities in the code, thereby significantly improving its vulnerability detection capabilities \cite{cao2024snopy}.

For the task of vulnerability detection, we aim to train a network $f_{e}(\cdot)$ that maps each graph $G_i \in \mathcal{G}$ to a low-dimensional vector $\mathbf{o}_i \in  \mathbb{R}^{d^{\prime}}$, and then use a classifier $f_{c}(\cdot)$ to perform binary classification on $\mathbf{o}_i$.

\subsubsection{Pipeline}
The overall architecture of \toolname is shown in Fig. \ref{model}, with its learning paradigm divided into four stages:

\begin{itemize}
\item \textbf{Parsing:} Given the source code of a function, we utilize static analysis tools to extract the code property graph of the function, which serves as the input to the following neural network.

\item \textbf{Hierarchical Graph Refinement:} This stage aims at reducing the complexity of the input graph, for which we propose an importance-based graph simplification module composed of several pooling layers (referred to as SAPool), each of which filtering a certain percentage of nodes from the input graph. Generally, according to the size of the code property graph, the inputs are self-adaptively fed into different numbers of SAPool layers to obtain the simplified code property graph.

\item \textbf{Context-aware Graph Encoding:} This stage aims at encoding the simplified graph and obtaining a vector representation for the final classification. To that end, we designed the global \& local feature aggregation module by integrating GNN with GT. Specifically, we design a cascading GNN and GT with residual connections as the backbone network, which encodes the refined code property graph from both global and local perspectives to obtain the graph representation. Subsequently, we employ global average pooling to embed the graph information into a vector.

\item \textbf{Classification:} In the final stage, we feed the graph embedding vector into a Multiple Layer Perception (MLP) layer to obtain an output probability, which denotes the prediction result, and train the network with the cross-entropy loss function.
\end{itemize}

Finally, during inference, given the source code of a function, \toolname can perform binary classification (vulnerable and non-vulnerable) in an end-to-end manner with the above well-trained models.
Please note that the two critical modules, \ie the hierarchical graph refinement and context-aware graph encoding are originally proposed in this study.

In the following subsections, we will detail the technical specifics of each stage.

\begin{figure}[!t]
\centering
\includegraphics[width=0.48\textwidth]{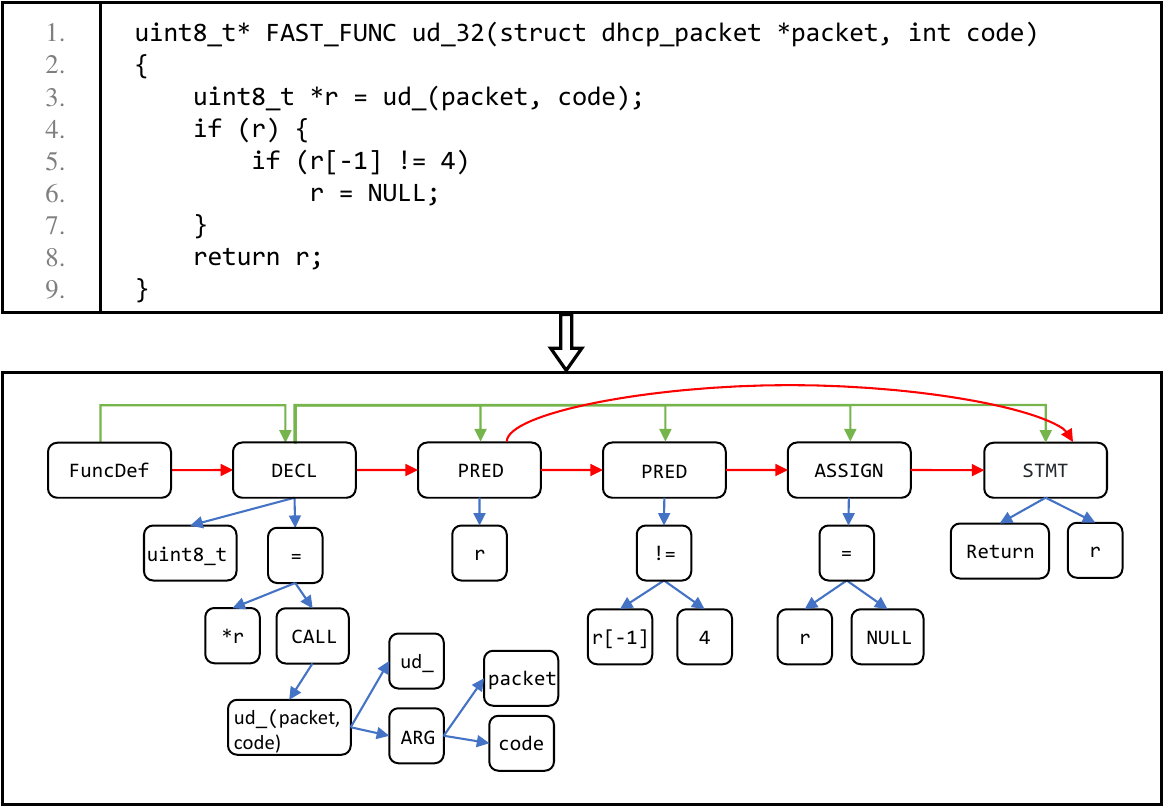}
\caption{A CPG for the example source code. The blue arrow indicates the AST structure, the red arrow indicates the CFG structure, and the green arrow indicates the PDG structure.}
\label{parse}
\end{figure}

\subsection{Parsing}
Code contains rich structural information, and using graphs can better represent the non-linear structure within the code. Similar to previous methods, \toolname utilizes the Joern tool to parse the source code, thereby generating a code property graph. The generated code property graph includes three important structures: AST, CFG, and PDG, which are rich in semantic and syntactic information. 
Specifically, given a source code file of a function, we first standardize the source code by removing comments from the code snippets. Next, we treat the source code as a natural language sequence to pre-train a Word2vec \cite{Arxiv2013word2vec} model. The clean source code is parsed into a code property graph ${G_i}$. For the edges in the graph, we convert the directed graph into an undirected graph. 
For nodes in the graph, the code contained in the node is treated as a token sequence, and its vector representation is initialized using the previously pre-trained Word2vec model. Within a single node, we take the average of all token vectors as the final attribute of that node.
Fig. \ref{parse} illustrates the corresponding code property graph of the first code snippet from our motivating examples, with the code graph structure subtly differentiated by the hues of the arrows.
Finally, we repeat the above process for all source files in the dataset to generate a set of graph ${\mathcal{G}}$.

\subsection{Hierarchical Graph Refinement}

In this stage, we introduce an importance-based graph simplification module to hierarchically reduce the size of the graph, where the nodes are filtered based on their importance scores. 
Considering that the input code property graphs are with various node sizes, we determine whether each code property graph requires undergoing the graph simplification based on a preset threshold. Specifically, if the size of the graph is below the threshold $T$, the graph will directly proceed to the following graph encoding stage; otherwise, it will go through a number of simplification steps. Such a process is self-adaptive, \ie it will automatically terminate when the graph size is below the threshold.

\subsubsection{The Pooling Operation in One Layer}
Pooling is the core operation to reduce the size of the graph. In this subsection, we introduce the pooling operation in \toolname, named as SAPool, whose workflow is divided into two steps: information aggregation and graph pooling.

The information aggregation step aims to gather structural and semantic information in large code graphs.

The prerequisite of this step is to accurately represent the information of each node. To that end, we employ a two-layer MLP as a projector to map the node information onto a higher-level embedding space, considering the feature extraction capability of MLP. To mitigate the issue of over-fitting during the training process, we also apply a dropout after the MLP. In this way, the model can more accurately capture the characteristics of the source code fragment contained in each node.

Specifically, given a code property graph $G = (\mathbf{X}, \mathbf{A})$ ($\mathbf{X}$ denotes the feature matrix, $\mathbf{A}$ denotes the adjacency matrix), this process can be represented as follows:
\begin{equation}
\mathbf{X} = \text{Dropout}(\mathbf{W}_{M_1}(\text{ReLU}(\mathbf{W}_{M_2}\mathbf{X}))),
\end{equation}
where $\mathbf{W}_{M_1}$ and $\mathbf{W}_{M_2}$ represent the learnable parameters in the MLP, and $\text{ReLU}(\cdot)$ denotes the activation function. In this process, the self-attribute information of nodes is enhanced.

The next step is information aggregation, aimed at perceiving the structural information within the code graph through the core message-passing operator of GNNs, thereby propagating the features contained between different nodes. Code graphs often exhibit complex structural relationships due to the presence of long-range paths and multiple branching structures. To capture the structural information of the code graph more accurately, we employ APPNP \cite{ICLR2018APPNP} (a well-known information propagation algorithm which denotes {\em Approximate Personalized Propagation of Neural Predictions}) to aggregate node information. We conducted comparative experiments on APPNP and other GNNs in Section \ref{sec:experimentalResult}. As a variant of GNNs, APPNP is better suited for learning code graphs compared to traditional GNNs (e.g., GGNN, GCN), because APPNP allows for adjustable message-passing ranges, facilitating the aggregation of information from distant graph nodes. In contrast, GNNs like GGNN and GCN focus solely on local neighborhood features, neglecting the long-range dependencies present in code graphs, which limits their effectiveness in tasks such as code vulnerability detection. Moreover, in SAPool, which focuses solely on large code graphs, APPNP does not require the introduction of additional training parameters, making it more efficient compared to traditional GNNs that introduce new parameters.

Specifically, APPNP executes information aggregation $l$ times in succession based on the graph structure. During each aggregation, it retains the node feature information with a probability of $\alpha$, and the neighbor feature information aggregated at the current layer is retained with a probability of $(1-\alpha)$. Finally, these two information resources are added together as the node's feature after this aggregation:
\begin{equation}
\label{appnp}
\mathbf{X}^{(l)}=(1-\alpha) \hat{\mathbf{A}}\mathbf{X}^{(l-1)}+\alpha \mathbf{X}^{(0)},
\end{equation}
where ${l}$ represents the number of times APPNP aggregates neighbor information, and ${\mathbf{X}^{(l)}}$ represents the output features of the APPNP layer. In this way, each node can aggregate neighbor information from ${1}$-th to ${l}$-th order, effectively capturing multi-scale contextual information within the graph.

After information aggregation, the model preliminarily extracts the features of the large graph. 
The core logic of a vulnerability typically involves only a few statements, whereas the entire function may consist of hundreds of lines of code. Other elements unrelated to the vulnerability act as noise, adversely affecting the performance of deep learning models \cite{cao2024snopy}. We employ Top-k pooling \cite{ICML2019TopK} to perform hierarchical pooling on large graphs. Compared to traditional global pooling operations (which generate a single one-dimensional vector), Top-k pooling filters redundant information by retaining subgraphs. We add a comparison between Top-k pooling and global pooling in subsequent experiments, which is shown in Section \ref{sec:experimentalResult}. The scoring mechanism of Top-k pooling is similar to the attention mechanism, which aims to help the model gradually pay attention to nodes associated with vulnerabilities, thereby simplifying the size of the code graph. This feature aids the model in filtering redundant information within code graphs, thereby enhancing the accuracy of feature extraction.

Next, we explain Top-k pooling in detail. Specifically, we use a learnable vector ${\mathbf{h}}$ to map each node feature ${\mathbf{X}}$ to a one-dimensional space as a score:
\begin{equation}
\mathbf{z} = \mathbf{X} \frac {\mathbf{h}}{||\mathbf{h}||},
\end{equation}
where ${||\cdot||}$ represents the L2 regularization. The score $\mathbf{z}$ determines the importance of each node. When the score is larger, it means that this node contains more important information. Next, we sort the nodes in descending order based on their scores and manually set a pooling rate $K$ as the proportion of nodes to retain. We select the important node indices according to $K$:
\begin{equation}
idx = \text{select}(\mathbf{z}, K), K \in (0, 1).
\end{equation}
Finally, the index is used to construct the adjacency and feature matrices of the simplified subgraph:
\begin{equation}
\mathbf{A} = \mathbf{A}[idx, idx],
\end{equation}
\begin{equation}
\mathbf{X} = \mathbf{X}[idx, :],
\end{equation}
\begin{equation}
\mathbf{z} = \mathbf{z}[idx],
\end{equation}
\begin{equation}
\mathbf{X} = \mathbf{X} \odot (\text{Sig} (\mathbf{z})),
\end{equation}
where $\text{Sig}(\cdot)$ denotes the sigmoid activation function. It is noted that the computation involves taking the Hadamard product of the projection values ${\mathbf{z}}$ and the feature matrix ${\mathbf{X}}$ of the selected top-k nodes. This process is crucial for the network to learn the weight vector $\mathbf{h}$ during back propagation. In this way, the importance of each node is automatically learned by the network according to the learned vector $\mathbf{h}$. Top-k pooling prompts the model to focus on the nodes that are most influential to the prediction, namely the parts of the code that are most likely to directly trigger vulnerabilities. The key information captured is crucial for vulnerability detection. Thus, the large graph is pooled into a smaller subgraph, containing only the selected important nodes and the edges between them, while many noise nodes in the large graph are filtered out.

\subsubsection{Self-Adaptive Pooling Numbers}

The above contents introduce the pooling operations in one pooling layer. It should be noted that merely pooling once may not be enough since the graph size might still be larger than the threshold after one SAPool (considering that there might be as many as several thousand nodes in the code graph). 
To address this challenge, \toolname adopts a self-adaptive strategy where the number of pooling layers is dynamically determined, \ie the pooling process is automatically terminated when the number of nodes in the graph is lower than the threshold $T$ (we determine the value of the hyperparameter $T$ in Section~\ref{Threshold}). This means that no matter how large the input code graph is, SAPool can simplify it to a code graph with fewer nodes than the threshold $T$.

Moreover, the pooling ratio of different layers is also dynamically adjusted. For the first SAPool layer, we set $K$ to 0.1, which means only the top ten percentage of the most important nodes are preserved, aiming at quickly filtering the irrelevant information at the beginning. While for the following layers, $K$ is increased by 0.1 each time, and its maximum value is set to 0.5.   
The behind intuition is that as the code graph simplifies, the proportion of important nodes gradually increases, and the important information might be lost if we maintain a low pooling ratio.

\subsection{Context-Aware Graph Encoding}
To capture the long-range dependencies in the graph, we designed the global \& local feature aggregation module consisting of multiple mixtures of GNN layer and GT layer as the backbone, with the detailed structure shown in Fig. \ref{model}. 
The GNN aims to aggregate local node features through message passing, while the GT aims to aggregate global node features through a multi-head self-attention mechanism.

\textbf{Local.} We employ a general GCN to aggregate information from neighbors. In a single layer GCN, only first-order neighbors can be aggregated. Therefore, the GCN aims to capture local dependencies within the graph structure. The operations in each GCN layer are as follows:
\begin{equation}
\mathbf{X} = \text{ReLU}({\hat{\mathbf{A}}\mathbf{X}\mathbf{W}_G}),
\end{equation}
where $\mathbf{W}_G$ denotes the learnable parameters in the GCN. Note that in the previous hierarchical graph refinement stage, the large graph may have been pooled into a non-connected graph. Solely using a GCN cannot extend information propagation to non-connected graphs and is limited in terms of long-distance features. To address this, we introduce GT to learn graph representations from a global perspective.

\textbf{Global.} We designed a GT based on multi-head self-attention mechanism. The fully connected learning mechanism inside has the following two advantages: (1) it captures long-distance dependencies between nodes; (2) it learns potential associations between subgraphs within non-connected graphs. The GT is divided into two steps: multi-head self-attention (MHA) and feed-forward blocks (FFN). First, in a single attention head, the attention score matrix ${\mathbf{A}}_{h}$ of each node is learned by all nodes, and the score matrix is utilized to update the node features:
\begin{equation}
{\mathbf{A}}_{h}=\text{softmax}\left(\frac{\left(\mathbf{X} {\mathbf{W}}_{Q}\right)\left(\mathbf{X}{\mathbf{W}}_{K}\right)^{\text{T}}}{\sqrt{d}}\right),
\end{equation}
\begin{equation}
\mathbf{X}_i = {\mathbf{A}}_{h} {\mathbf{X}} {\mathbf{W}}_{V},
\end{equation}
where ${\mathbf{W}}_{Q}$, ${\mathbf{W}}_{K}$, and ${\mathbf{W}}_{V}$ are the learnable parameter matrices in the MHA. The outputs X from different heads are concatenated to obtain the final output of MHA:
\begin{equation}
\mathbf{X} = \text{Concat} (\mathbf{X}_1, \mathbf{X}_2,...,\mathbf{X}_i).
\end{equation}
Through this method, the global attention in the graph is captured, and the information of each node is updated by the features from all the other nodes in the graph. Next, the output from MHA is fed into the FFN, and residual connections are added respectively in MHA and FFN:
\begin{equation}
\mathbf{X} = \text{MHA}(\mathbf{X}) + \mathbf{X},
\end{equation}
\begin{equation}
\mathbf{X} = \text{FFN}(\mathbf{X}) + \mathbf{X}.
\end{equation}
The FFN consists of a two-layer MLP that expands and then compresses dimensions:
\begin{equation}
\mathbf{X} = \mathbf{W}_{F1}(\text{LN}(\mathbf{X})),
\end{equation}
\begin{equation}
\mathbf{X} = \mathbf{W}_{F2}(\text{LN}(\mathbf{X})),
\end{equation}
where $\text{LN}(\cdot)$ represents layer normalization \cite{Arxiv2016LayerNorm}, $\mathbf{W}_{F1}$ and $\mathbf{W}_{F2}$ represents the learnable parameters in the FFN.

We add residual connections twice within the GNN-GT framework to enhance the propagation of feature information. Thanks to the complementary feature propagation of GNN and GT layers, both global and local information are adequately exploited to facilitate the extraction of vulnerability information in code graphs.

Finally, we compress the node representation obtained by backbone network encoding into a one-dimensional vector for classification based on global average:
\begin{equation}
\mathbf{o} = \text{AvgPool}(\mathbf{X}).
\end{equation}

\subsection{Classification}
In the final stage of the paradigm, we feed the previously obtained one-dimensional vector $\mathbf{x}$ into a two-layer MLP, and then directly output the probabilities of two categories (vulnerable and non-vulnerable). We use binary cross-entropy loss to estimate the loss between the predicted labels and the actual labels: 
\begin{equation}
\mathcal{L} = -\frac{1}{M} \sum_{i=1}^{M} [\mathbf{y}_i \log(\mathbf{p}_i) + (1 - \mathbf{y}_i) \log(1 - \mathbf{p}_i)],
\end{equation}
where $\mathcal{L}$ is the loss, $M$ is the number of graphs, $\mathbf{y}_i$ is the actual label of the $i$-th graph, and $\mathbf{p}_i$ is the predicted probability of the $i$-th graph as a vulnerable code snippet. 
During the training process of \toolname, we perform forward propagation to calculate the loss and update the learnable parameters in the model through backward propagation.

\section{Experimental Setup}
\label{sec:experimentalSetup}
\subsection{Research Questions}
To verify the effectiveness of \toolname, we aim to answer the following research questions:

\textbf{RQ1: Effectiveness.} Can \toolname surpass current state-of-the-art learning-based vulnerability detection methods?

\textbf{RQ2: Rationale.} To what extent do different design choices affect the overall performance of \toolname?

\textbf{RQ3: Sensitivity.} To what extent do different hyper-parameters affect the overall performance of \toolname?

\subsection{Experimental Setup}
\subsubsection{Datasets}

\begin{table}
\centering
\refstepcounter{table}
\label{dataset}
\caption{Statistics of datasets.}
\begin{tabular}{c!{\vrule width \lightrulewidth}c!{\vrule width \lightrulewidth}c!{\vrule width \lightrulewidth}c!{\vrule width \lightrulewidth}c} 
\toprule
Dataset & Vulnerable & Non-vulnerable & Total  & Language  \\ 
\midrule
Devign \cite{Nips2019Devign}  & 10067         & 12294             & 22361  & C         \\
Reveal \cite{TSE2021Reveal} & 1658          & 16511             & 18169  & C/C++     \\
Big-Vul \cite{MSR2020BigVul} & 10900         & 177736            & 188636 & C/C++     \\
\bottomrule
\end{tabular}
\end{table}
We evaluate the effectiveness of the method using three widely used public vulnerability datasets, namely Devign \cite{Nips2019Devign}, Reveal \cite{TSE2021Reveal}, and Big-Vul \cite{MSR2020BigVul}. The distribution information of the three datasets is shown in Table \ref{dataset}. The Devign dataset consists of two manually labeled real-world projects, FFMPeg and Qemu. It contains a total of about 22k samples, with the proportion of vulnerable samples being 45.02\%. The Reveal dataset comes from two open-source projects, Debian and Chromium, and contains a total of about 18k samples, with the vulnerability proportion being 10.04\%. Big-Vul is collected from a large number of GitHub projects, including 91 different types of vulnerabilities, with a vulnerability proportion of 5.77\%.

\subsubsection{Baseline Methods}
In our experiment, we compared \toolname with night state-of-the-art vulnerability detection methods, two based on tokens while the other six based on graphs. Detailed descriptions of the baseline methods used for comparison are provided below.

\textbf{(1) Flawfinder} \cite{lipp2022empirical}: Flawfinder is a static analysis tool that integrates rules to identify C/C++ source code functions that are susceptible to vulnerabilities.

\textbf{(2) VulDeePecker} \cite{Arxiv2018Vuldeepecker}: VulDeePecker utilizes a bidirectional LSTM to learn the natural language sequence representation of code for vulnerability detection.

\textbf{(3) SySeVR} \cite{TDSC2021Sysevr}: SySeVR utilizes a bidirectional RNN to learn the program representation of vulnerability-related syntactic and semantic information.

\textbf{(4) Devign} \cite{Nips2019Devign}: This method combines multiple graph structures as input and uses GGNN and CNN to extract features for vulnerability detection. 

\textbf{(5) Reveal} \cite{TSE2021Reveal}: Reveal consists of a feature extraction and training phase. It first parses the code into a Code Property Graph, then builds the model using GRU and GGNN. 

\textbf{(6) IVDetect} \cite{FSE2021IVDetect}: IVDetect takes the PDG as input and uses GCN for vulnerability detection. 

\textbf{(7) DeepWuKong} \cite{TOSEM2021Deepwukong}: DeepWuKong extracts semantic features of the code through program slicing methods and uses GNN to learn structured information for vulnerability detection. 

\textbf{(8) MAGNET} \cite{TSE2023MAGNET}: MAGNET views the code structure graph as a heterogeneous graph and uses a meta-path-based attention graph neural network to extract features.

\textbf{(9) AMPLE} \cite{ICSE2023AMPLE}: AMPLE is based on manually defined rules to reduce the size of the code structure graph and uses multi-head attention mechanisms and CNN to learn general graph representations. 

Please note that the baselines are all open sourced, and we directly reuse their implementations in this study to avoid potential threats in the reproduction process.

\subsubsection{Evaluation Metrics}
To comprehensively assess the performance of our proposed method, we employ the following standard evaluation metrics: Accuracy, Precision, Recall, and the F1 score. Each metric offers insight into different aspects of the model's prediction capabilities.

\textbf{Precision: }$\text{Precision} = \frac{\text{TP}}{\text{TP + FP}}$. Precision measures the ratio of correctly predicted positive observations to the total predicted positive observations. High precision relates to a low rate of false positives. 

\textbf{Recall: }$\text{Recall} = \frac{\text{TP}}{\text{TP + FN}}$. Recall measures the ratio of correctly predicted positive observations to all observations in actual class. Recall emphasizes the model's ability to find all relevant instances in the dataset.

\textbf{F1 score: }$\text{F1} = 2 \times \frac{\text{Precision} \times \text{Recall}}{\text{Precision} + \text{Recall}}$. F1 score is the weighted average of precision and recall. Therefore, this score takes both false positives and false negatives into account. It is especially useful in the case of uneven class distribution.

\textbf{Accuracy:} $\text{Accuracy} = \frac{\text{TP + TN}}{\text{TP + TN + FN + FP}}$. Accuracy is the most intuitive performance measure and it is simply the ratio of correctly predicted instances to the total instances in the dataset.

\subsection{\toolname Implementation}
\label{sec:implementation}
We implement \toolname using the Pytorch and PyTorch Geometric libraries in Python 3.10. The model is trained on a single NVIDIA GeForce RTX 4090 GPU. For the initial dimension of Word2vec, we follow the setting of the previous method \cite{ICSE2023AMPLE} and set it to 100. In terms of parameter settings for the model, we mainly focused on the threshold $T$ (10, 20, 30, 40, 50, 60, 70, 80, 90, 100), aggregation times $l$ (2, 4, 6, 8, 10), retention probability $\alpha$ (0.1, 0.2, 0.4, 0.6, 0.8), GNN-GT layers (2, 3, 4, 5, 6, 7, 8), head numbers (1, 2, 4, 8, 16). Finally, the threshold $T$, aggregation times $l$, retention probability $\alpha$, GNN-GT layers and head numbers are empirically determined to 40, 8, 0.2, 5, 4, respectively. Moreover, the hidden dimension of the model is set to 64. During the model training phase, the batch size is set to 1024, with a maximum iteration count of 3000. AdamW \cite{loshchilov2018AdamW} is used as the optimizer, with a learning rate of 1e-3 and weight decay of 1e-5. We replicate all baseline models based on publicly released source codes and use the same hyper-parameter settings as described in their original papers. Finally, to ensure a fair comparison, all methods are subjected to the same partitioning standard. The datasets are randomly divided into training, validation, and testing sets in a 7:1:2 ratio.

This experimental setup aligns with that of the Reveal study \cite{TSE2021Reveal}, which aims to enhance the testing dataset size to evaluate the generalizability of the approaches. 
Please note that the original AMPLE study splits the dataset based on a 8:1:1 ratio \cite{ICSE2023AMPLE}.
Consequently, there exists a disparity between the outcomes acquired in our study and those initially reported.
Specifically, the initially reported performance of AMPLE slightly surpasses the results obtained in our experiments (as will be demonstrated in Section~\ref{sec:experimentalResult}), possibly due to the availability of more data for comprehensive training.

\section{Experimental Result}
\label{sec:experimentalResult}

\begin{table*}
\centering
\caption{Comparison results between \toolname and the baselines. The \textbf{boldface} and \underline{underline} values indicate the best and the runner-up results (\%), respectively.}
\label{SOTA}
\scriptsize
\setlength{\tabcolsep}{4pt}

\begin{tabular}{c|l|cccc|cccc|cccc} 
\toprule
\multirow{2}{*}{Type}   & Dataset      & \multicolumn{4}{c|}{Devign}                                       & \multicolumn{4}{c|}{Reveal}                                       & \multicolumn{4}{c}{Big-Vul}                                        \\ 
\cmidrule{2-14}
                             & Method       & Accuracy       & Precision      & Recall         & F1 score       & Accuracy       & Precision      & Recall         & F1 score       & Accuracy       & Precision      & Recall         & F1 score        \\ 
\cmidrule{1-1}\cmidrule{2-14}
Static Analysis & Flawfinder & \underline{54.00}          & 47.31       & 14.67               & 22.39          & 79.89          &    14.79        & 21.01          & 17.36          & 88.72          & 12.45           & 16.53          & 14.20           \\
\cmidrule{1-1}\cmidrule{2-14}
\multirow{2}{*}{Token-based} & VulDeePecker & 46.46          & 44.78          & 42.73          & 43.73          & 69.49          & 8.87           & 23.13          & 12.82          & 74.76          & 6.56           & 25.29          & 10.42           \\
                             & SySeVR       & 49.44          & 43.56          & 39.11          & 41.21          & 77.47          & 13.50          & 23.50          & 17.15          & 77.84          & 9.32           & 32.95          & 14.53           \\ 
\cmidrule{1-1}\cmidrule{2-14}
\multirow{6}{*}{Graph-based} & Devign       & 48.13          & 45.11          & 49.10          & 47.02          & 76.71          & 11.30          & 19.00          & 14.17          & 79.73          & 9.10           & 24.14          & 13.22           \\
                             & DeepWuKong   & 50.19          & 46.40          & 49.92          & 48.09          & 81.25          & 18.67          & 25.42          & 21.53          & 77.74          & 9.81           & 30.31          & 14.83           \\
                             & Reveal       & 49.85          & 46.78          & 50.70          & 48.66          & 81.30          & 18.18          & 24.23          & 20.77          & 82.81          & 13.38          & 30.84          & 18.66           \\
                             & IVDetect     & 50.66          & 47.36          & 47.25          & 47.30          & 79.93          & 16.88          & 24.82          & 20.10          & 79.37          & 11.15          & 31.95          & 16.53           \\
                             & MAGNET       & 51.43          & 48.55          & \underline{60.77}  & \underline{53.98}  & 82.17          & 23.82          & 34.68          & 28.24          & 81.22          & 12.96          & 33.90          & 18.75           \\
                             & AMPLE        & 53.12  & \underline{50.00}  & 55.92          & 52.80          & \underline{82.79}  & \underline{26.99}  & \textbf{41.09} & \underline{32.58}  & \underline{83.60}  & \underline{15.77}  & \underline{36.06}  & \underline{21.95}   \\ 
\cmidrule{1-1}\cmidrule{2-14}
Graph-based                  & \toolname         & \textbf{58.87} & \textbf{55.08} & \textbf{66.41} & \textbf{60.22} & \textbf{87.45} & \textbf{37.77} & \underline{37.05}  & \textbf{37.41} & \textbf{89.47} & \textbf{26.96} & \textbf{37.85} & \textbf{31.49}  \\
\bottomrule
\end{tabular}
\end{table*}

\subsection{Comparison with State-of-the-Arts (RQ1)}
To address the first research question, we compared \toolname with eight state-of-the-art baseline methods across three datasets. Subsequently, we visualized the latent variables in the context-aware graph encoding stage to further validate the effectiveness of \toolname.
\subsubsection{Comparison of Results}
The comparison analysis in Table \ref{SOTA} demonstrates the overall performance results of \toolname across four evaluation metrics. Such comparisons have led us to some key insights. First, \toolname consistently surpasses the proposed baseline methods in ACC and F1 scores across all three datasets, reflecting its superior effectiveness. 
Compared with the static analysis method Flawfinder, \toolname shows comprehensive advantages. Taking the Devign dataset as an example, the F1 score of Flawfinder is only 22.39\%, with a high false negative rate.
Compared to token-based methods, \toolname shows significant improvements across four evaluation metrics on three datasets. Specifically, \toolname achieves an absolute improvement of 14.71\% in ACC and 21.07\% in F1 over the token-based approach VulDeePecker on the Big-Vul dataset. This is because token-based methods treat code as simple natural language sequences, overlooking structural information. In contrast, graph-based methods can better capture structural information in code, which is advantageous for vulnerability detection. 
Despite this, \toolname still outperforms the existing graph-based approaches. Compared with the best-performing graph-based method AMPLE on the Big-Vul dataset, \toolname achieves an absolute increase of 5.87\% in ACC and 9.54\% in F1 score. Although AMPLE reduces the number of nodes to some extent by using artificially defined rules, the proportion of simplification based on these rules might be less than 20\% (according to our manual investigations). As a comparison, our \toolname can reduce the size of the graph by more than 70\% on average (as will be discussed in Section~\ref{sec:discussion}).
In code graphs, AMPLE still faces the problem of the large amount of noise and limited understanding ability towards long-range dependencies, resulting in the suboptimal outcomes.
When compared with the most trivial graph-based baseline method Devign in terms of F1 on three datasets, \toolname achieves an absolute increase of 13.2\%, 23.24\%, and 18.27\%. This is because Reveal merely transforms the code into a graph structure and then trains the GNN model merely based on the original code graphs, thus facing greater challenges when the code graphs are complex.
Furthermore, when considering all the 12 settings (4 performance metrics $\times$ 3 datasets), \toolname achieves the best performances in 11 cases, with the remaining one (recall on the Reveal dataset) being the second-highest. 

\begin{figure}[!t]
\centering
\includegraphics[width=0.5\textwidth]{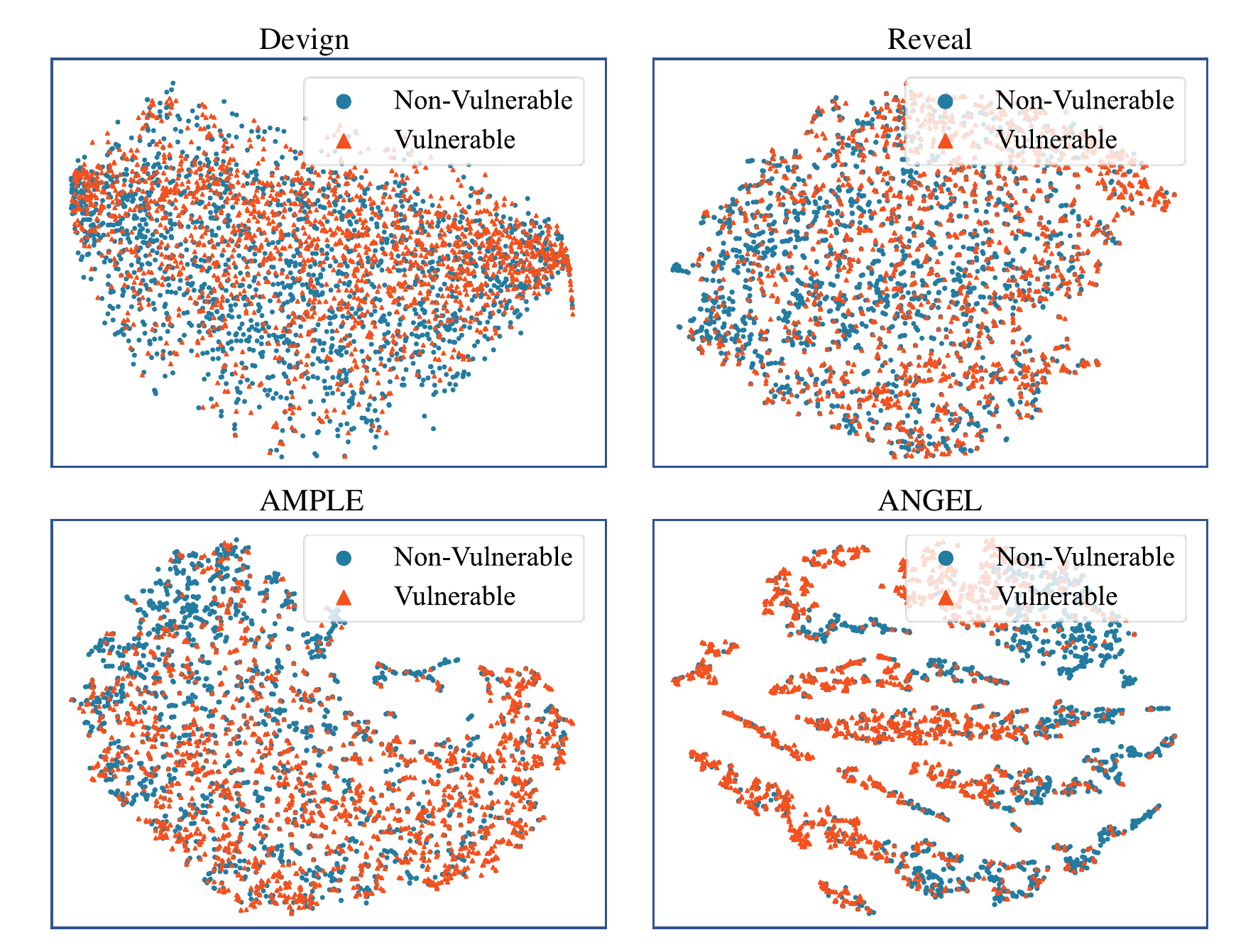}
\caption{The T-SNE algorithm \cite{JMLR2008T-SNE} visualization illustrates the distribution between vulnerable and non-vulnerable samples in the code representations of different methods.}
\label{T-SNE}
\end{figure}

\subsubsection{Visualization of Results}
To directly observe the effectiveness of \toolname, we used the T-SNE algorithm \cite{JMLR2008T-SNE} to visualize the graph-level embeddings, helping us understand how different categories of data are distinguished in the latent space. 
We selected four baseline methods for embedding visualization on the Devign dataset. The experimental results are shown in Fig. \ref{T-SNE}, where red dots represent vulnerable samples and blue dots represent non-vulnerable samples. When using the T-SNE algorithm for embedding visualization, it is usually expected that data of different categories form relatively independent cluster centers in the low-dimensional space. Based on Fig. \ref{T-SNE}, we make the following observations: (1) The Devign method entangles the two categories of data. This indicates that Devign fails to adequately capture the features distinguishing the different categories, making it difficult to separate them in the latent space;  (2) The performance of Reveal is similar to that of Devign, as it also confuses the two categories of data in the embedding space;  (3) AMPLE performs better than the previous two methods, with more non-vulnerable samples clustering in the top-left corner and more vulnerable samples clustering in the bottom-right corner. This suggests that AMPLE has stronger feature extraction capabilities for both types of samples. However, overall, AMPLE does not distinctly separate the two categories of samples;  (4) \toolname shows obvious separability under visual observation, where the vulnerable samples are mainly concentrated on the left, and the non-vulnerable samples are mainly concentrated on the right. Relative to the three baseline methods, the embeddings generated by \toolname are more distinguishable in the latent space. Such results demonstrate the effectiveness of \toolname in distinguishing vulnerable and non-vulnerable codes.

\begin{tcolorbox}
\textbf{Answer to RQ1:} \toolname exhibits improved effectiveness, performing better than all baseline methods in terms of ACC and F1 scores. Compared to the best-performing baseline AMPLE, \toolname achieves a 14.0\%-43.4\% increase in terms of the F1 scores across three datasets. Visualization further indicates that \toolname captures key information in the code more effectively than baseline methods.
\end{tcolorbox}

\begin{table*}
\centering
\caption{Results of ablation study.}
\label{RQ2}
\begin{tabular}{l|cccc|cccc|cccc} 
\toprule
\multirow{2}{*}{Module} & \multicolumn{4}{c|}{Devign}                                       & \multicolumn{4}{c|}{Reveal}                                       & \multicolumn{4}{c}{Big-Vul}                                        \\ 
\cmidrule{2-13}
                        & Accuracy       & Precision      & Recall         & F1 score       & Accuracy       & Precision      & Recall         & F1 score       & Accuracy       & Precision      & Recall         & F1 score        \\ 
\midrule
``W/O HGR'' &55.82 & 52.91 & 52.38 & 52.64 & 82.87 & 30.63 & 31.35 & 30.99 & 83.42 & 19.25 & 26.15 & 22.18  \\
``W/O GNN'' &57.96 & 54.48 & 62.71 & 58.31 & 86.36 & 36.64 & 34.20 & 35.38 & 88.54 & 23.81 & 36.00 & 28.67  \\
``W/O GT''  &56.75 & 53.32 & 62.05 & 57.35 & 85.21 & 35.66 & 32.78 & 34.16 & 86.63 & 24.51 & 29.94 & 26.96  \\
\midrule
\toolname & \textbf{58.87} & \textbf{55.08} & \textbf{66.41} & \textbf{60.22} & \textbf{87.45} & \textbf{37.77} & \textbf{37.05}  & \textbf{37.41} & \textbf{89.47} & \textbf{26.96} & \textbf{37.85} & \textbf{31.49}  \\
\bottomrule
\end{tabular}
\end{table*}
\subsection{Ablation Analysis (RQ2)}
\subsubsection{The impact of different modules}
To answer the second research question, we conducted ablation studies on three datasets to explore the impact of each component in \toolname on the results. We compare three different variants of \toolname: (1) without hierarchical graph refinement (denoted as ``w/o HGR''); (2) without local aggregation in GNN (denoted as ``w/o GNN''); (3) without global aggregation in GT (denoted as ``w/o GT'').
The experimental results are shown in Table \ref{RQ2}, with the performance of all the variants being lower than \toolname, indicating that every component contributes to the overall performance of \toolname. 
Specifically, several key observations were summarized: (1) For the variant ``w/o HGR", the accuracy, precision, recall, and F1 scores decline by an average of 4.56\%, 5.67\%, 10.48\%, and 7.77\% across the three datasets, respectively, which are the most significant impacts on the overall performance. This suggests that hierarchical graph refinement plays a crucial role in vulnerability detection, helping the models concentrate on critical information; (2) For the variant ``w/o GNN'', there was a decrease of 0.98\%, 1.63\%, 2.80\%, and 2.25\% respectively in the four metrics, indicating that GNNs are effective in aggregating neighbor information during the feature extraction process; (3) Compared to the variant ``w/o GT'', \toolname showed improvements of 2.40\%, 2.11\%, 5.51\%, and 3.55\% respectively in the four metrics. This demonstrates that GT can overcome the limitations of GNNs, extracting potential global dependencies in the code structures.

\begin{table}
\centering
\caption{The impact of different GNNs on SAPool.}
\label{RQ3-R1-appnp}
\begin{tabular}{c|c|c|c}
\toprule
Method  & Devign & Reveal & Big-Vul \\ 
\midrule
GCN & 53.61  & 33.56  & 29.06    \\
GGNN  & 57.49  & 36.27  & 29.70    \\
GAT  & 55.24  & 34.37  & 30.46    \\
APPNP  & 60.22  & 37.41  & 31.49    \\
\bottomrule
\end{tabular}
\end{table}

\subsubsection{The Impact of APPNP in SAPool}

In SAPool, we introduce an algorithm called APPNP, designed to perform information aggregation in code graphs. We select three classic GNNs (\ie GGNN \cite{Arxiv2015GGNN}, GCN \cite{ICLR2017GCN}, and GAT \cite{ICLR2018GAT}) as variants to replace APPNP in order to observe the changes in detection performance. Table \ref{RQ3-R1-appnp} presents the F1 scores of different variants across three datasets, and based on these, we observe the following: (1) GCN is simple, efficient, and has low computational complexity, but it yields poor results when applied to SAPool, with F1 scores of 53.61\%, 33.56\%, and 29.06\% on the three datasets, respectively; (2)GGNN and GAT dynamically propagate and aggregate information from neighboring nodes through gated recurrent units and attention mechanisms, respectively, achieving superior results compared to GCN; (3) When APPNP is used, the model achieves the best results, with F1 scores of 60.22\%, 37.41\%, and 31.49\% on the three datasets, respectively. This is because GCN, GGNN, and GAT focus solely on learning from first-order neighboring nodes, failing to capture the structural relationships between higher-order neighbors. This suggests that relying exclusively on these traditional GNNs makes it challenging to extract long-range dependency features, \ie control flow and data flow, from code graphs. In contrast, APPNP possesses the ability to aggregate long-distance features, enabling the model to better learn the feature within the code graph.

\begin{table}
\centering
\caption{The impact of different pooling strategies on SAPool.}
\label{RQ3-R1-topk}
\begin{tabular}{c|c|c|c}
\toprule
Method  & Devign & Reveal & Big-Vul \\ 
\midrule
Avg  & 47.57  & 28.57  & 24.53   \\
Sum  & 48.68  & 27.03  & 22.92    \\
Max  & 42.64  & 21.05  & 20.60    \\
Top-k  & 60.22  & 37.41  & 31.49    \\
\bottomrule
\end{tabular}
\end{table}

\subsubsection{The Impact of Top-k in SAPool}

To validate the effectiveness of the Top-k pooling strategy, we select three different global pooling methods as baselines: global average pooling (denoted as ``Avg"), global sum pooling (denoted as ``Sum"), and global max pooling (denoted as ``Max"). ``Avg" and ``Sum" compute a one-dimensional vector by averaging or summing all node features, respectively. ``Max" retains the maximum feature value among all nodes as a new one-dimensional vector. Table \ref{RQ3-R1-topk} illustrates the F1 scores of different pooling strategies across three datasets. It is evident that the detection performance of the model under the three global pooling baselines is significantly lower than that of Top-k pooling. Specifically, both ``Avg'' and ``Sum'' force all semantic and structural information of nodes in the graph to be compressed into a one-dimensional vector, causing confusion in node information and thus resulting in poor detection performance. ``Max" considers only the maximum value of each feature dimension, resulting in a significant loss of node information in the pooled vector, which leads to the worst performance among the various baselines. Compared to previous global pooling methods, Top-k pooling is more suitable for learning large code graphs. During each pooling operation, it compresses the code graph into a smaller subgraph, preserving pivotal node information relevant to the categories. Through a hierarchical pooling process, large code graphs are simplified into smaller subgraphs, making it easier for subsequent graph encoding models to extract the features of the code graph.

\begin{tcolorbox}
\textbf{Answer to RQ2:} Each component of \toolname effectively brings positive gains to the model, highlighting the rationale of the proposed components. Among them, hierarchical graph refinement contributes the most to the overall performance.
 \end{tcolorbox}

\begin{table}
\centering
\caption{The F1 scores under different thresholds $T$ in the importance-based graph simplification module.}
\label{RQ3-1}
\begin{tabular}{c|c|c|c} 
\toprule
\diagbox{Threshold $T$}{Dataset} & Devign         & Reveal         & Big-Vul         \\ 
\midrule
10                                 & 53.39          & 30.92          & 29.66           \\
20                                 & 57.43          & 33.75          & 29.99           \\
30                                 & 59.05          & 37.06          & 30.38           \\
40                                 & \textbf{60.22} & \textbf{37.41} & \textbf{31.49}  \\
50                                 & 59.22          & 36.56          & 30.15           \\
60                                 & 57.15          & 35.82          & 29.28           \\
70                                 & 55.40          & 35.52          & 28.96           \\
80                                 & 53.83          & 33.95          & 26.28           \\
90                                 & 52.60          & 30.70          & 25.54           \\
100                                & 51.55          & 29.14          & 24.13           \\
\bottomrule
\end{tabular}
\end{table}

\subsection{Hyper-Parameter Sensitivity Analysis (RQ3)}
In this section, we explore the impact of hyper-parameters on the performance of \toolname, including settings of the pooling ratio of SAPool, hyper-parameter configurations of APPNP, and the number of layers and attention heads in GNN-GT.

\subsubsection{The Impact of Threshold in the Importance-based Graph Simplification Module}
\label{Threshold}
We explore the impact of different threshold $T$ in the importance-based graph simplification module on \toolname across three datasets. According to Table \ref{RQ3-1}, we can observe that: (1) When the threshold is set between 30 and 60, \toolname achieves better performance; (2) When the threshold is set too small, F1 score of \toolname will decrease. This is because retaining too few nodes may not be able to involve enough vulnerability-related information, leading to the loss of important information in the graph; (3) When the threshold is set too high, the code graph is not sufficiently simplified, making it difficult for the model to learn the vulnerability pattern. Overall, based on observations from the three datasets, setting the threshold $T$ to 40 allows \toolname to achieve the optimal results.

\begin{figure}[!t]
\centering
\includegraphics[width=0.49\textwidth]{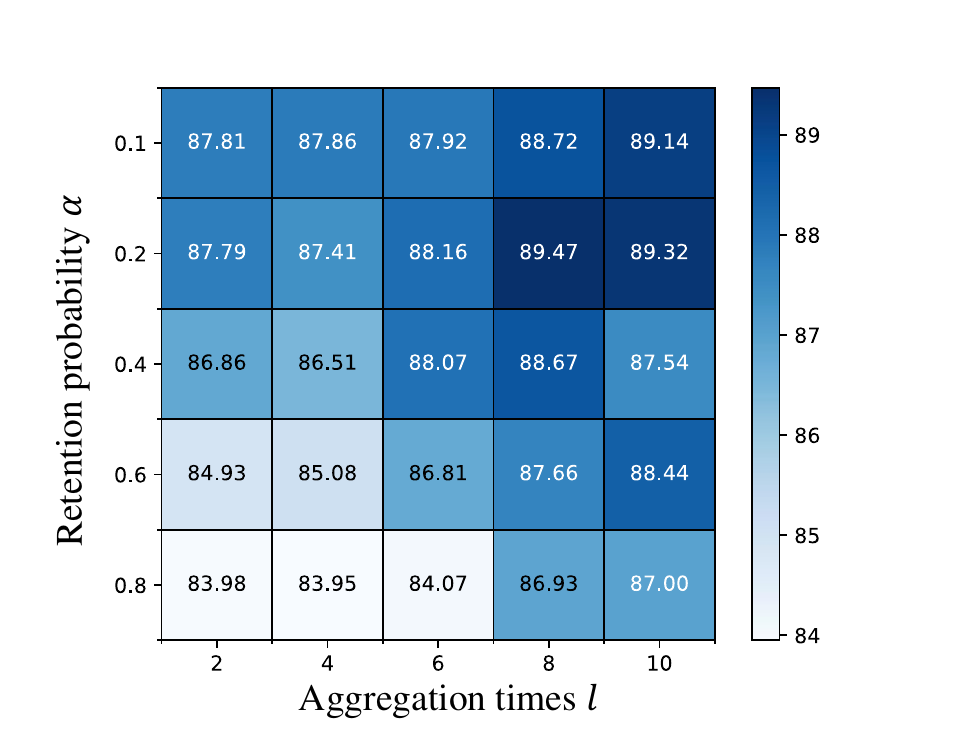}
\caption{Accuracy under different parameter settings in the APPNP layer on the Big-Vul \cite{MSR2020BigVul} dataset.}
\label{APPNPparameters}
\end{figure}

\subsubsection{The Impact of Preset Parameters in APPNP}
Eq. (\ref{appnp}) introduces two hyper-parameters, $l$ and $\alpha$, to influence the distance of message propagation and the importance of the node information. Fig. \ref{APPNPparameters} shows the effects of hyper-parameters $l$ and $\alpha$ on the network. Specifically, we can observe the following phenomenons: (1) The performance of \toolname increases with the increase of $l$. When $l$ reaches 8, the growth trend tends to flatten; (2) \toolname achieves superior performance when the $\alpha$ is small. In general, \toolname can better capture the structural information in the graph when the propagation distance $l$ and the node information weight $\alpha$ are 8 and 0.2 respectively.

\begin{figure}[!t]
\centering
\includegraphics[width=0.49\textwidth]{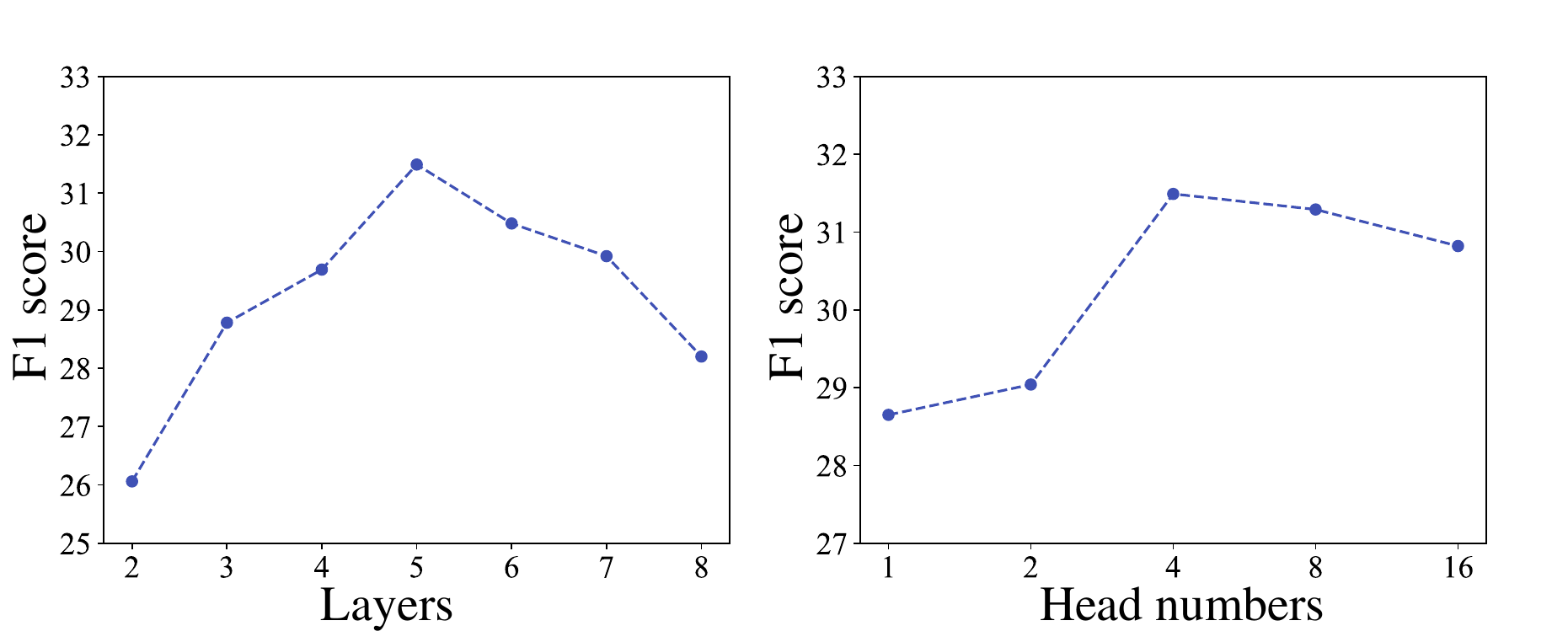}
\caption{Parameter analysis of GNN-GT layers and head numbers of GT on the Big-Vul \cite{MSR2020BigVul} dataset.}
\label{RQ3-3}
\end{figure}

\begin{table}
\centering
\caption{The F1 score of different pooling rates $K$ in Top-k pooling.}
\label{RQ3-R1}
\begin{tabular}{c|c|c|c}
\toprule
\diagbox{Pooling rate $K$}{Dataset} & Devign & Reveal & Big-Vul  \\ 
\midrule
Fixed to 0.1                          & 56.20  & 34.21  & 28.80    \\
Fixed to 0.5                          & 59.94  & 37.10  & 31.68    \\
Increment from 0.1 to 0.5             & 60.22  & 37.41  & 31.49    \\
\bottomrule
\end{tabular}
\end{table}

\subsubsection{The Impact of pooling rate in SAPool}

We investigate the impact of different pooling rates in SAPool on \toolname across three datasets. We set three different pooling rates: fixed at 0.1, fixed at 0.5, and incrementally increasing from 0.1 to 0.5. As observed in Table \ref{RQ3-R1}, the following conclusions can be drawn: (1) When the pooling rate in SAPool is consistently set to 0.1, \toolname performs the worst. This is due to the fact that maintaining such a small pooling rate continuously compresses the code graph into overly small subgraphs, leading to significant information loss; (2) When the pooling rate is fixed at 0.5, a significant performance improvement is observed compared to 0.1, achieving performance comparable to that of incremental pooling rates. However, this fixed pooling rate of 0.5 leads to an increase in the number of adaptive pooling operations, resulting in an increase in model parameters and a reduction in efficiency. This is because ANGLE needs to ensure that the number of nodes in the simplified code graph remains below the threshold $T$; (3) As the pooling rate gradually increases from 0.1 to 0.5, the model achieves optimal performance on both datasets (\ie Reveal and Big-Vul) while having fewer pooling layers than when $K$ is fixed at 0.5. Therefore, we set the pooling rate to an increasing sequence from 0.1 to 0.5.

\subsubsection{The Number of Layers and The Number of Attention Heads in GNN-GT}
We analyzed the impact of different numbers of layers and attention heads in GNN-GT on the performance of \toolname. As shown in Fig. \ref{RQ3-3}, on the Big-Vul dataset, both the number of layers and the number of attention heads in GNN-GT affect the performance of \toolname. We observed that \toolname achieves optimal performance when the number of GNN-GT layers is set to 5 and the number of attention heads is also set to 4. When the number of layers and attention heads in GNN-GT is too large, the performance of \toolname shows a declining trend, which is due to the over-fitting of the model.

\begin{tcolorbox}
\textbf{Answer to RQ3:} The setting of hyper-parameters affects the performance of \toolname across multiple datasets. To simplify the parameter configuration, we set the same hyper-parameters on different datasets while achieving relatively good performance.
\end{tcolorbox}

\section{Discussion}
\label{sec:discussion}

\begin{table}
\centering
\caption{The accuracy of \toolname and variants in different node number ranges on the Big-Vul \cite{MSR2020BigVul} dataset.}
\label{Dis}
\begin{tabular}{l|c|c|c} 
\toprule
Method  & $(0, 25]$ & $(25, 50]$ & $(50, 100]$ \\ 
\midrule
``w/o HGR" & 95.51 & 93.15 & 84.49\\
\toolname & 96.89 & 95.23  & 91.66 \\
\bottomrule
\end{tabular}
\end{table}

\begin{figure}[!t]
\centering
\includegraphics[width=0.48\textwidth]{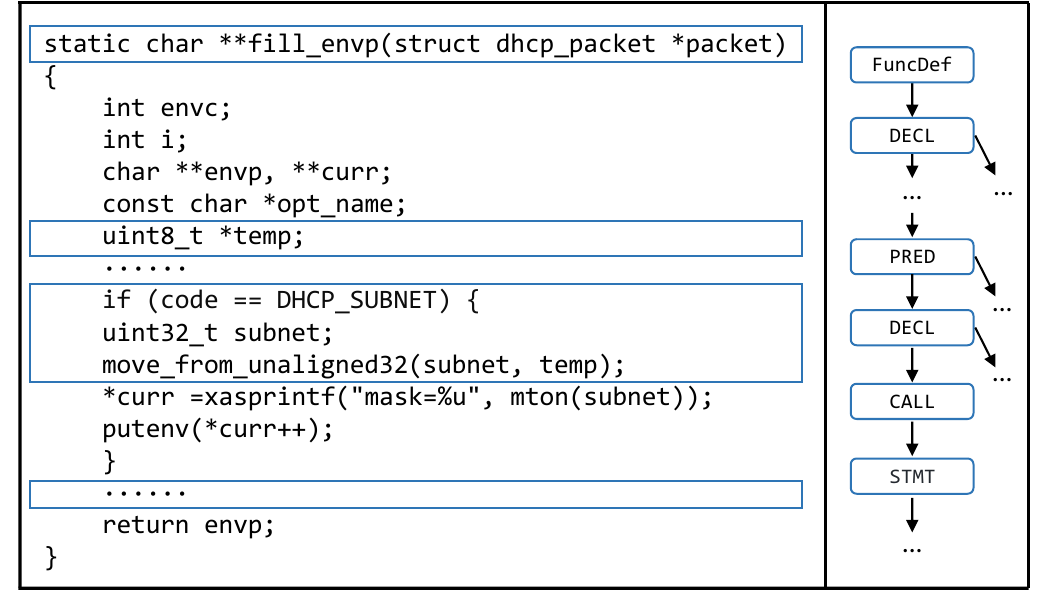}
\caption{An example of graph simplification. The blue boxes represent the retained statements after simplification.}
\label{simpli}
\end{figure}

\subsection{Why Dose \toolname Work?}
\subsubsection{Advantages of ANGLE}
We identify that \toolname has two advantages that can explain its effectiveness in source code vulnerability detection.

\textbf{(1) The ability for adaptive simplification of code graphs.} The proposed hierarchical graph refinement stage simplifies graphs with a large number of nodes into subgraphs with critical nodes, helping \toolname more effectively capture key information in large graphs. 
Specifically, we calculate the average number of nodes in the input and output of the hierarchical graph refinement stage.
Taking the Big-Vul \cite{MSR2020BigVul} dataset as an example, the number of nodes in the simplified graph decreases by an average of 75.9\% compared to the original input, from 104 to 25.
Additionally, we explored the impact of the hierarchical graph refinement stage on samples with different numbers of nodes in the Big-Vul dataset. Similarly, the variant model configuration was ``w/o HGR", and the experimental results are shown in Table \ref{Dis}. We found that when the number of nodes is (0, 25], the gap between \toolname and ``w/o HGR" is only 1.38\%. As the node count interval increases, the gap between \toolname and ``w/o HGR" also gradually increases. In the range where the number of nodes exceeds 50, the performance gap increases from 1.38\% to 7.17\%. This indicates that hierarchical graph refinement can effectively mine critical information in large graphs, thereby making the model generalize well to complex code graphs. 
An example is shown in Fig.~\ref{simpli} where we demonstrate the simplified code graph of the second code snippet from our motivating examples in Section~\ref{sec:motivation}. The initial graph comprises over 400 nodes, whereas in the simplified version, it condenses to just 21 nodes.
Nevertheless, as depicted in the graph, the majority of retained nodes are vulnerability-relevant, such as those associated with the declaration and invocation of the {\mycode temp} variable, which leads to the vulnerability.

\textbf{(2) The capability to perceive local and global contexts information.} We have designed a module that aggregates both global and local features, aiming at learning informative content within the graph. It utilizes GNN to aggregate local information and employs GT to aggregate global information. 
The GT assists the GNN in capturing long-distance dependencies and avoiding the over-smoothing problem. GNN helps the GT to enhance local information, preventing important neighbor features from being overwhelmed by global features.
As demonstrated in Table \ref{RQ2}, when GNN and GT are used independently, the detection performance of model decreases in both cases. Such results suggest that there is a complementarity between GNN and GT. Therefore, the GNN-GT layer that we designed by combining GNN and GT effectively captures the structural relationships within the graph, thereby enhancing the performance of vulnerability detection.

\begin{figure}[!t]
\centering
\includegraphics[width=0.48\textwidth]{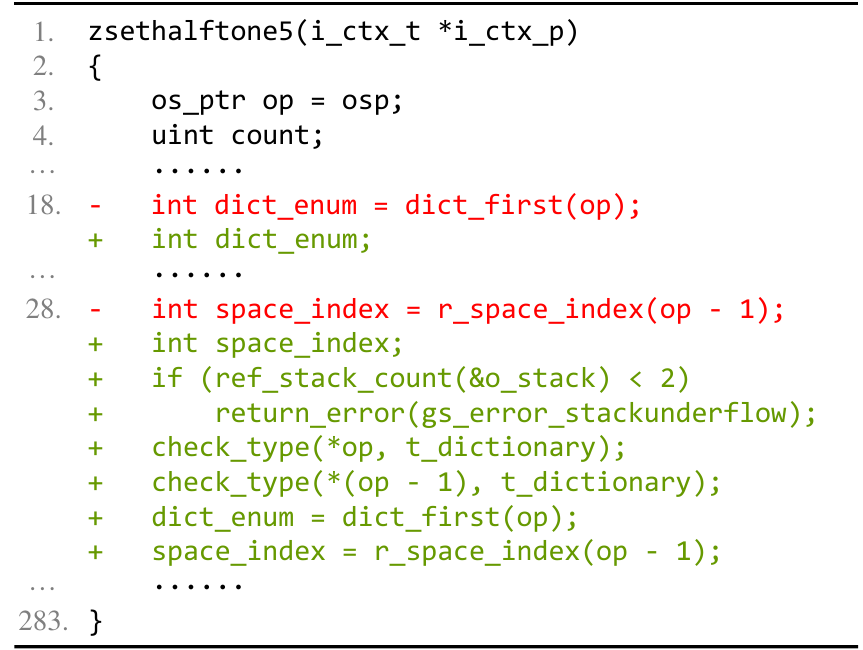}
\caption{A source code of CWE-704 example. The red-colored code is vulnerable code, and the green-colored code is the repaired code.}
\label{Limitate}
\end{figure}

\begin{figure}[!t]
\centering
\includegraphics[width=0.49\textwidth]{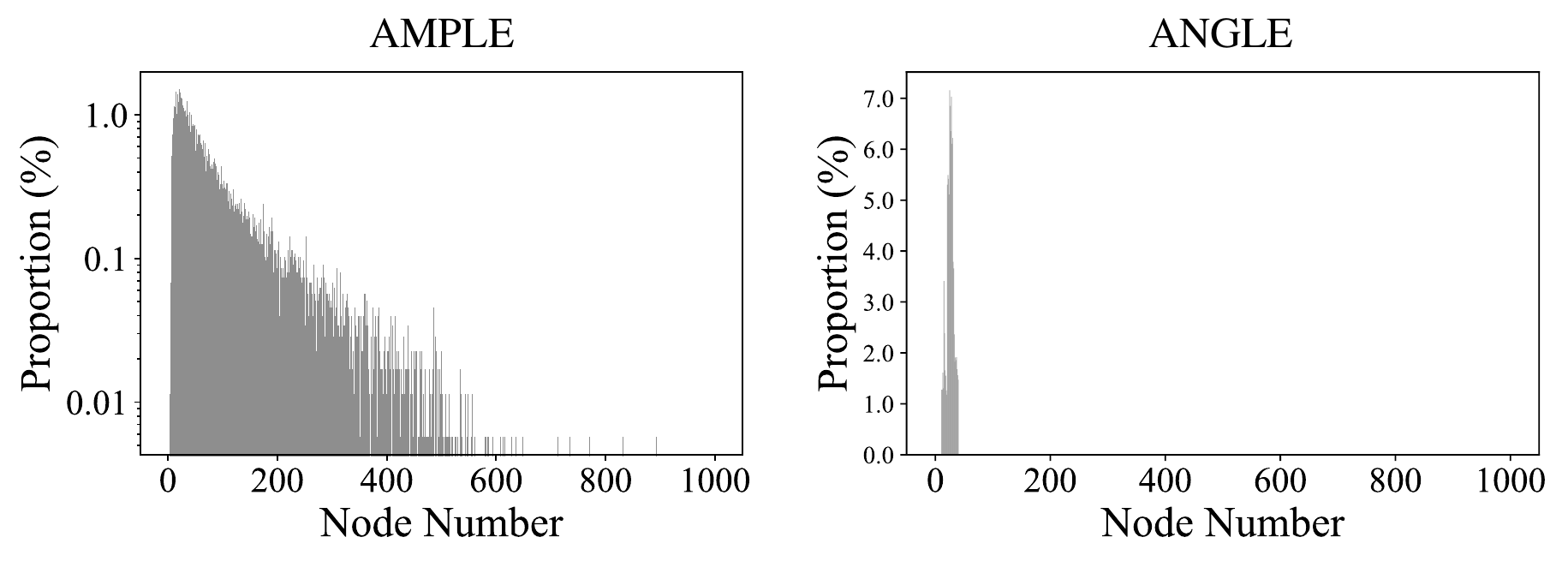}
\caption{The statistical distribution of the sizes of simplified code graphs for AMPLE and \toolname on the Reveal dataset.}
\label{AMPLEafter}
\end{figure}

\begin{table}
\centering
\caption{Comparison of the average number of nodes after simplification between \toolname and AMPLE in different node number ranges.}

\label{Disnodenum}
\begin{tabular}{c|c|c|c}
\toprule
Interval  &  Original  & AMPLE & \toolname \\ 
\midrule
$[$100, 200)  & 143  & 76  & 29    \\
$[$200, 300) & 244  & 125  & 25    \\
$[$300, 1000)  & 511  & 254  & 30    \\

\bottomrule
\end{tabular}
\end{table}

\subsubsection{The Impact of Different Code Graph Sizes on Detection Results} 

In Fig. \ref{NodeAcc}, we present the results of ANGLE and the state-of-the-art method AMPLE on code graphs of varying sizes. Next, we discuss the reasons behind the performance differences between AMPLE and ANGLE on large graphs. AMPLE removes nodes with redundant semantics during the data pre-processing phase based on manually predefined rules. However, due to the limitations of these rules, this approach can only simplify a small number of redundant nodes. As shown in Fig. \ref{AMPLEafter}, we illustrate the size of the simplified code property graphs of AMPLE and \toolname on the Reveal dataset. We can visually observe that AMPLE still contains a significant number of samples with more than 100 nodes in the simplified code graphs, while in contrast, ANGLE reduces all the code graphs to fewer than 40 nodes. Our statistics reveal that the average node count for AMPLE in the simplified code graphs is 91, whereas ANGLE's average node count is only 26. Overall, ANGLE demonstrates a significantly superior ability to compress graph size compared to AMPLE. This difference leads to overall performance of AMPLE being inferior to ANGLE across multiple datasets and evaluation metrics. Furthermore, as shown in Fig. \ref{NodeAcc}, we observe a noticeable performance gap between AMPLE and ANGLE when the node count exceeds 100. We conducted further analysis on the compression of code graph sizes by AMPLE and ANGLE when the node count exceeds 100. As illustrated in Table \ref{Disnodenum}, AMPLE, which is based on a rule-based graph simplification method, faces limitations when handling large code graphs. Specifically, as the original code graph size increases, AMPLE's simplification performance gradually diminishes. When the node count exceeds 300, the average node count in AMPLE’s simplified graphs remains as high as 254. In contrast, ANGLE employs a hierarchical graph refinement strategy, which can effectively adapt to code graphs of varying sizes, resulting in superior detection performance. Thus, the performance gap between AMPLE and ANGLE grows as the node count increases.

\subsection{Limitation}
Despite \toolname achieving superior performance over baseline methods in vulnerability detection tasks, its F1 scores on the Reveal \cite{TSE2021Reveal} and Big-Vul \cite{MSR2020BigVul} datasets are only 37.4\% and 31.5\%, respectively. The reason for this phenomenon is the extreme imbalance between vulnerable and non-vulnerable samples in the datasets, with different types of vulnerabilities within the vulnerable samples following a long-tail distribution. 
Specifically, taking the Big-Vul dataset as an example, CWE-119 accounts for 19.94\% of vulnerable samples, while CWE-22 accounts for only 0.33\%. This imbalanced distribution makes it difficult for the model to adequately learn the vulnerability patterns of the minority samples, leading to incorrect predictions for these samples.
As shown in Fig. \ref{Limitate}, the examples of CWE-704 demonstrate issues with incorrect type conversion transformations. First, we observe that the vulnerability pattern is complex, with a large number of program elements being involved in the bug fix. 
Furthermore, vulnerabilities with the type CWE-704 account for only 0.22\% in the Big-Vul dataset, making the vulnerability pattern of this instance challenging for the model to mine. 
Therefore, these vulnerabilities are misclassified by the model due to their scarce sample numbers and the uneven distribution of data.
Our future work will explore how to generate more vulnerability samples within datasets and introduce resampling techniques during the training phase to further aid the model in enhancing its performance with imbalanced samples.

\begin{table}
\centering
\caption{Comparison of resource consumption between \toolname and the baselines.}
\label{ResourceConsumption}
\begin{tabular}{l|c|c|c} 
\toprule
Method  & \multicolumn{1}{c|}{Parameters (M)}& \multicolumn{1}{c|}{Time Consumption (s)}& \multicolumn{1}{c}{F1 Score} \\ 
\midrule
Devign & 0.76 & 11.7 & 13.22 \\
Reveal & 1.07 & 13.8 & 18.66 \\
AMPLE  & 1.73 & 110.5 & 21.95 \\
\toolname   & 0.30 & 24.2 & 31.49\\
\bottomrule
\end{tabular}
\end{table}
\subsection{Resource Consumption}
When exploring the performance of a model, resource consumption is an indispensable factor to consider. 
Resource consumption mainly includes two aspects: model size and time consumption, both of which are related to the versatility. 
In this section, we will further analyze the efficiency of \toolname by comparing its performance with baseline models in these two aspects. Model size refers to the total number of parameters in the whole model.
Time consumption refers to the time spent by the model to infer a test set. 

It is important to highlight that in this study, all the baselines are reproduced on our server, which means their running environment is identical to that of \toolname, as detailed in Section~\ref{sec:implementation}.
This practice guarantees a fair comparison across various approaches in terms of the time consumption.

According to Table \ref{ResourceConsumption}, we can have the following observations:

\begin{itemize}
    \item \toolname has the least amount of parameters in the model, which is because the model architecture we designed follows the principle of ``narrow while deep'' \cite{ICML2017expressive}, \ie the hidden dimension is small while the model is deep. Typically, existing methods \cite{Nips2019Devign, TSE2021Reveal, ICSE2023AMPLE} set the hidden layer dimension to 200, while \toolname has a hidden layer dimension of only 64. Our preliminary experiments show that directly increasing the hidden layer dimension of \toolname does not result in a significant performance gain in the results. Therefore, our approach is more lightweight compared to existing methods.
    \item There are some non-parameter computing operations in \toolname, such as APPNP and matrix multiplication in the attention mechanism. Therefore, although \toolname has the least amount of parameters, it is at a medium level in terms of time spent.
    \item Relying solely on stacking model parameters, the model design of Devign and Reveal is too simplified to capture vulnerability patterns well. Therefore, although they have advantages in terms of time spent, their effectiveness is decreased significantly compared with \toolname.
\end{itemize}
Such results indicate that \toolname is not only effective in terms of the vulnerability detection task but also lightweight with respect to the model size, which makes it easy to be deployed and executed.

\subsection{Threats to Validity}
\subsubsection{External Threats}
All our results and findings are limited to three publicly available benchmark datasets. However, these datasets were created with the C/C++ language. Other programming languages such as Python and Java have not yet been implemented. In principle, \toolname can be applicable to other programming languages, as the method does not rely on specific artificial rules for its implementation. In the future, we will attempt to select a greater variety of programming language datasets to evaluate our approach.

\subsubsection{Internal Threats}
During the data pre-processing phase and the model training phase, there are some uncertainties. Specifically, we may be unable to precisely standardize the code in the data pre-processing phase, and we use the tool Joern to parse the source code. To our knowledge, Joern is frequently updated. Different versions of Joern have slight differences in the parsing of source code, which leads to uncertainties in the input for the model. Additionally, during the model training phase, we rely on random grid search to determine the optimal hyper-parameters as best as we can, but it is impossible to try all combinations.


\section{Conclusion}
\label{sec:conclusion}
In this paper, we reveal the bottleneck of current vulnerability detectors that they usually fail to predict accurately on large code graphs. We further propose \toolname, a hierarchical and context-aware graph representation learning approach for vulnerability detection. 
\toolname is designed to (1) hierarchically reduce the size of the graph to alleviate the interference of noise information, and (2) utilize the complementarity of GT and GNN to understand the semantics of the code graph from both the global and local perspectives. 
Experimental results demonstrate that \toolname is effective in vulnerability detection, achieving higher performance compared to state-of-the-art methods and mitigating the ineffectiveness of existing vulnerability detection techniques on large code graphs.

\ifCLASSOPTIONcaptionsoff
  \newpage
\fi

\bibliographystyle{IEEEtran}
\bibliography{cite}

\end{document}